%% file: end_state_arxiv_draft3.tex
\documentclass[english,british]{agujournal2019}
\usepackage[LGR,T1]{fontenc}
\usepackage{textcomp}
\usepackage[utf8]{inputenc}
\usepackage{verbatim}
\usepackage{amsmath}
\usepackage{amssymb}
\usepackage{graphicx}
\usepackage{rotfloat}
\usepackage[authoryear]{natbib}
\usepackage{microtype}
\AtBeginDocument{%
  \justifying
  \hyphenpenalty=10000     
  \exhyphenpenalty=10000   
  \tolerance=9999          
  \emergencystretch=3em    
}

\makeatletter

\usepackage{txfonts}
\usepackage{textgreek}
\usepackage{graphicx}
\usepackage{aas-macros}
\usepackage{setspace}
\usepackage{layout}
\journalname{}

\makeatother

\usepackage{babel}
\begin{document}

\title{The climate and potential habitability of a warm and oxygenated
Mars}

\authors{Ashwin S. Braude, \affil{\selectlanguage{english}%
1}\\
Edwin S. Kite, \affil{\selectlanguage{english}%
1,2}\\
Preston Cosslett Kemeny, \affil{\selectlanguage{english}%
3}\\
Raymond T. Pierrehumbert, \affil{\selectlanguage{english}%
4,5}\\
Alexandre Kling, \affil{\selectlanguage{english}%
1}\\
and Michael A. Mischna, \affil{\selectlanguage{english}%
6}}

\affiliation{1}{Astera Institute, Emeryville, CA, USA.}

\affiliation{2}{University of Chicago, Chicago,
IL, USA (kite@uchicago.edu).}

\affiliation{3}{Woods Hole Oceanographic Institution, Woods Hole, MA, USA.}

\affiliation{4}{MIT, Cambridge, MA, USA.}

\affiliation{5}{University of Oxford, Oxford, UK.}

\affiliation{6}{Jet Propulsion Laboratory, California Institute of Technology, Pasadena,
CA, USA.}

\begin{abstract}
We present the first 3-D Global Climate Model simulations of an oxygen-rich, terraformed Mars. Assuming an O\textsubscript{2}--CO\textsubscript{2} atmosphere, with additional warming from a well-mixed grey gas, we study (1)~how surface temperatures change with atmospheric pressure and grey gas opacity, and (2)~how temperature, pressure, and other effects limit life as oxygen builds up. We neglect the climate effects of dust and the water cycle, as well as possible soil-chemistry and nutrient limits to life. We find that (1) increasing grey gas opacity and increasing atmospheric pressure both moderate day--night temperature swings, but  only slightly reduce seasonal temperature swings. Climate remains harsher in the Southern mid-latitudes relative to the Northern hemisphere. (2) For an atmospheric pressure of 30~hPa (5~times its present-day value), temperatures on Mars would be consistent with lichen and moss, followed at 60~hPa by forests (initially in the Hellas Basin). However, creating a human-breathable atmosphere through photosynthesis alone, at the oxygen production rates of biomes on Earth today, would take at least thousands of years, and require more oxygen than can be made via photosynthesis of known CO\textsubscript{2} reserves on Mars alone (unless CO\textsubscript{2} is recycled).
\end{abstract}

\section{Introduction}

Mars today is a hostile environment for life. While the search for microbes on Mars continues, it is plausible that in the future emphasis might shift to sustaining human life and ecosystems. The idea of making Mars' climate suitable for complex life has been studied for over 50 years (e.g., \citealt{sagan1973,averner1976,mckay1991,marinova2005,dicaire2013,kite2026roadmap}). Mars has: a) a past climate that was, at least intermittently, warm and wet (e.g., \citealt{craddockhoward2002,hynek2010,kiteconway2024,mondro2025}), b) large reserves of water ice \citep{jakoskyhallis2024}, and c) many of the elemental building blocks of life, including carbon, sulphur \citep{franz2019}, nitrogen in the form of nitrates \citep{Stern2015}, and phosphorus likely in the form of phosphates \citep{hausrath2024}. However, Mars has an eccentric orbit (with perihelion during southern summer) and more intense seasons than on Earth, despite a similar axial tilt. A necessary, but insufficient, first step to allow life to grow \citep{2020AsBio..20..163M,cockell2024} would be warming to allow for liquid water. Warming could employ solid-state greenhouse effect blankets at small scale \citep{wordsworth2019}; at regional-to-global scale, options include an artificial greenhouse effect \citep{marinova2005,dicaire2013,richardson2025}. 

Even with warming, however, Mars' average atmospheric pressure at the surface (6~hPa, 95\%~CO\textsubscript{2}) is close to the triple point of water. At these pressures, surface liquid water would undergo extreme evaporitic cooling, while even small pressure changes could cause boiling. The temperature range allowing surface liquid is narrow \citep{langeforget2026}, restricting liquid water to the shallow subsurface. The atmosphere is also too thin to damp the \textasciitilde 100 K day--night surface temperature fluctuations \citep{Atri2023}. Together with the high level of UV irradiation, and salty, perchlorate-rich soil, this makes the growth of surface life challenging (e.g., \citealt{schuergernicholson2006,berry2010effects}). Nonetheless, microbes might in principle be engineered to tolerate these stresses and grow where liquid water is available (e.g., \citealt{CaroAstorga2024, martin2026}). Adding Earth-derived microbes to Mars soils would not occur until after much more intensive searches for pre-existing life on Mars (e.g., \citealt{kitewordsworth2025, sephton2025}). Extremophile `pioneer' species might then release O\textsubscript{2}---currently very scarce in the Mars atmosphere (\textasciitilde 1 Pa, e.g., \citealt{sandel2015})---for example, through photosynthesis \citep{debenedictis2025}. The resulting O\textsubscript{2}-enriched atmosphere might support oxygen-requiring life \citep{mckaymarinova2001}.

What kind of biosphere could be sustained on Mars depends on how much O\textsubscript{2} can be built up. Even if the known buried CO\textsubscript{2} ice and adsorbed CO\textsubscript{2}  reserves on Mars were outgassed by warming, this would only give  \textasciitilde20~hPa of CO\textsubscript{2} \citep{jakoskyedwards2018} with a corresponding upper limit of  at most 20~hPa of photosynthetic O\textsubscript{2}. This level of oxygen in the atmosphere would not allow humans to breathe unaided and would make plant growth difficult  (see \ref{sec:Minimal-conditions-for}). Moreover, some O\textsubscript{2} would be lost to oxidative weathering and aerobic respiration (as for the Biosphere 2 experiment, \citealt{dreamingthebiosphere}). To get more oxygen, options include a) abiotic water-splitting or b) biological reduction of oxidised species in the soil, with sequestration of the resulting reductants to prevent re-oxidation. It is uncertain whether these mechanisms (\ref{sec:How-much-oxygen}) could release enough oxygen into the Martian atmosphere to allow flowers and forests. It is also unclear if a nitrogen cycle---needed for a complex biosphere---could be sustained without the Haber process or nitrogen imports, given Mars' low partial pressure of atmospheric nitrogen. This study emphasises climate, with a more detailed biogeochemical analysis left for future work. 

\citet{graham2004} proposed an ecological succession (`ecosynthesis') based on mountain ecosystems on Earth. This begins with cyanobacteria, followed by lichens, mosses, and more complex plants, to scale up oxygen production and bootstrap the ecosystem. On Earth, this process occurred over geologic time \citep{knollnowak2017} and, without active assistance, it would be very slow on Mars as well. Using a 1-D surface energy balance model, \citet{wendland2026} showed that an artificially warmed 100~hPa CO\textsubscript{2}--O\textsubscript{2} atmosphere on Mars would have temperatures allowing pine forests (\emph{Pinus sibirica}). However, to allow complex life, the climate must moderate day--night and seasonal temperature swings. This cannot be accurately represented using a system of connected 1-D~models, which use eddy diffusion to approximate horizontal advection (e.g., \citealt{charnay2020}).

Here, for the first time, we use a 3-D Global Climate Model (GCM) to simulate a terraformed Mars with a CO\textsubscript{2}--O\textsubscript{2} atmosphere. We vary oxygen build-up and artificial greenhouse warming to~map out potentially habitable climates. In Section \ref{sec:Model} we describe our climate model and ecological assumptions. In Section \ref{sec:Results} we map out temperature stressors across Mars as a function of a) atmospheric thickness and b) atmospheric heating. Using temperature and oxygen limits for~10~species including humans, we map out the potential for life to grow on Mars as terraforming proceeds, and estimate how long oxygenation would take. We focus on the Hellas Basin, a large crater in the Southern hemisphere with the planet's lowest elevations and highest atmospheric pressures. We~discuss model limitations and future work in Section \ref{sec:Conclusions-and-future}, and present conclusions in Section \ref{sec:Conclusions}.

\section{Model}\label{sec:Model}

We simulate Mars surface temperatures using the MarsWRF 3-D GCM \citep{michalakes2004,Skamarock2005,richardson2007,toigo2012}. 
MarsWRF solves the equations of atmospheric dynamics on a global grid, as well as the radiative balance of incoming shortwave radiation and outgoing longwave radiation, and heat and momentum transport through advection and turbulent~mixing. The model also exchanges energy and momentum with the surface and shallow subsurface. We run the model at 60 $\times$ 36 longitude--latitude spatial resolution and with 52 vertical layers from the surface to an altitude of 90 km in modified-sigma coordinates. The surface topography is smoothed using a 5-point nearest-neighbour smoother to minimise dynamical instabilities around sharp topographic features. Radiative transfer is modelled in 15 spectral bins in the~visible and near-infrared (0.2~-~5 $\mu$m) and 25 spectral bins in the mid- to thermal infrared (5~-~1000 $\mu$m), subdivided logarithmically in wavenumber space. The model shows <4\% net radiation imbalance at the top of the atmosphere averaged over the Mars year (see Table~\ref{fluxleakstable}), with the radiative balance understating outgoing longwave flux for high grey gas forcing values and overstating outgoing longwave flux for low grey gas forcing values. This causes a slight overestimate in~the difference in global average surface temperatures between our lowest and highest grey gas forcing values, without changing our conclusions.

We present 24 GCM simulations, all with a fixed partial pressure of CO\textsubscript{2} (18~hPa). We consider six pressure values of radiatively inert O\textsubscript{2} (quantified in terms of combined (CO\textsubscript{2} + O\textsubscript{2}) pressure at Mars datum $P_{dat}=$ 3, 5, 10, 15, 20 and 30 times the present-day Mars value of $P_{dat} =~$6.1~hPa). For each pressure value, we consider four opacity levels, set relative to the total CO\textsubscript{2} and O\textsubscript{2} partial pressures, of an idealised `grey gas' that is radiatively active but has negligible partial pressure ($k$ = 0, $\frac{2.5}{P_{dat}}$, $\frac{5.0}{P_{dat}}$ and $\frac{10.0}{P_{dat}}$~m\textsuperscript{2}~kg\textsuperscript{-1}). Our~lowest-pressure simulations assume a 97\% CO\textsubscript{2} atmosphere with a datum pressure ($P_{dat}$) of~18.3~hPa (3~times that of modern-day Mars), which corresponds approximately to the outgassing of known polar CO\textsubscript{2} ice deposits \citep{buhler2020} and adsorbed CO\textsubscript{2} \citep{jakoskyedwards2018} in the soil, plus the \textasciitilde 0.3~hPa \citep{2019JGRE..124.3000T} of radiatively inert gases currently present in the Mars atmosphere. We assume that the rest of the atmosphere in each simulation is O\textsubscript{2}, which pressure-broadens CO\textsubscript{2} absorption but is otherwise radiatively inert. The mean molecular weight of the atmosphere is calculated as a CO\textsubscript{2}--O\textsubscript{2} mixture. In practice, producing O\textsubscript{2} by photosynthesis would reduce the partial pressure of CO\textsubscript{2}. This is also necessary for humans, who cannot tolerate CO\textsubscript{2} levels above 3~hPa \citep{nioshco2}. Nonetheless, it is convenient for this study to keep CO\textsubscript{2} partial pressures constant in order to decouple the effects of CO\textsubscript{2} absorption from broadening by O\textsubscript{2} on Mars climate. This choice has little effect, as for a given total pressure, varying the O\textsubscript{2} fraction changes global average surface temperatures by <5 K (Fig.~\ref{co2o2mix}). Upper limits on atmospheric pressure would be tied to the practical challenges of oxygenation (see \ref{sec:How-much-oxygen} for further discussion).

The grey gas opacity is a proxy for an artificial warming agent deployed on Mars to~add to the natural CO\textsubscript{2} greenhouse. This could be a super-greenhouse gas (e.g., \citealt{gerstell2001,marinova2005}), an engineered aerosol \citep{ansari2024,richardson2025}, or a system of orbital reflectors \citep{Essunfeld2026}. The grey gas is assumed to be well-mixed in the atmosphere, and is transparent to incoming solar radiation between 0.2 and 5~$\mu$m, and absorbs outgoing longwave radiation at a constant opacity between 5 and 1000 $\mu$m.

We assume a quasi-infinite north polar cap of water ice %
that remains intact even with warming. Our simulations were benchmarked against modern Mars in \citet{braude2026arxiv} by tuning the polar CO\textsubscript{2} ice cap albedos and emissivities to fit the annual cycle of atmospheric pressure \citep{Guo2009}. The thermal inertia of the surface is from \citet{putzigmellon2007} and we neglect rises in soil thermal inertia with higher atmospheric pressure \citep{mischnapiqueux2020} which would help to moderate temperature swings.
We neglect any effects of the water cycle on~the~climate system, including rain, rivers, and the radiative effects
of clouds \citep{braude2026arxiv}. Modelling these would require better knowledge of cloud microphysics
on a warmed Mars (see Section \ref{sec:Conclusions-and-future}), while greenhouse warming from water vapour alone is likely to be small \citep{braude2026arxiv}. 
We therefore only take into account temperature constraints of drought-resistant
organisms in this study. We also exclude all effects of atmospheric dust on
the climate system. Warming might increase the amount of dust in Mars' atmosphere, which would cool daytime surface temperatures. However, we do not understand Mars' dust
cycle well enough to predict how it would change on a warm Mars (see \citealt{kahre2017} for a review). Organisms might suppress dust lifting by binding the soil, as on Earth \citep{wu2020trade}.

We considered ten species to explore ecological succession \citep{graham2004}. A real-world succession would have to take into account not just climate constraints, but also the composition of the soil, and would be informed by advances in synthetic biology, a field that is evolving too rapidly \citep{martin2026,li2026} to be forecast here. The first species may be a cyanobacterium that can survive exposure to Mars-like conditions \citep{devera2019}, that can photosynthesise in anoxic conditions while surviving in, for instance, ice-covered lakes that buffer against temperature cycles \citep{kling2020}. As oxygen increases, first lichen \citep{devera2010,devera2014} and then desert moss \citep{Li2024} can be introduced. Increased oxygen build-up and~radiation protection might then allow for more plants, including radiation-resistant flowering plants that can tolerate Antarctic conditions \citep{graham2004}, and an altitude-tolerant pine tree \citep{wendland2026} that can survive extreme cold. 

At this stage, humans might introduce familiar food crops, though the atmosphere
would still have too little oxygen for humans to breathe. We consider a selection of food crops---sea buckthorn, wheat, kale and lentils---based on a) their temperature tolerance and b) their ability to~provide a~balanced diet. We also map out where humans can live on Mars without a~pressure suit but with supplementary oxygen, and where humans can live on Mars and breathe Mars air. More detail on~each organism's habitability constraints is given in Tables \ref{organismtable1}-\ref{organismtables2} and \ref{sec:Minimal-conditions-for}.

\begin{figure}
\centering
\includegraphics[width=0.75\textwidth]{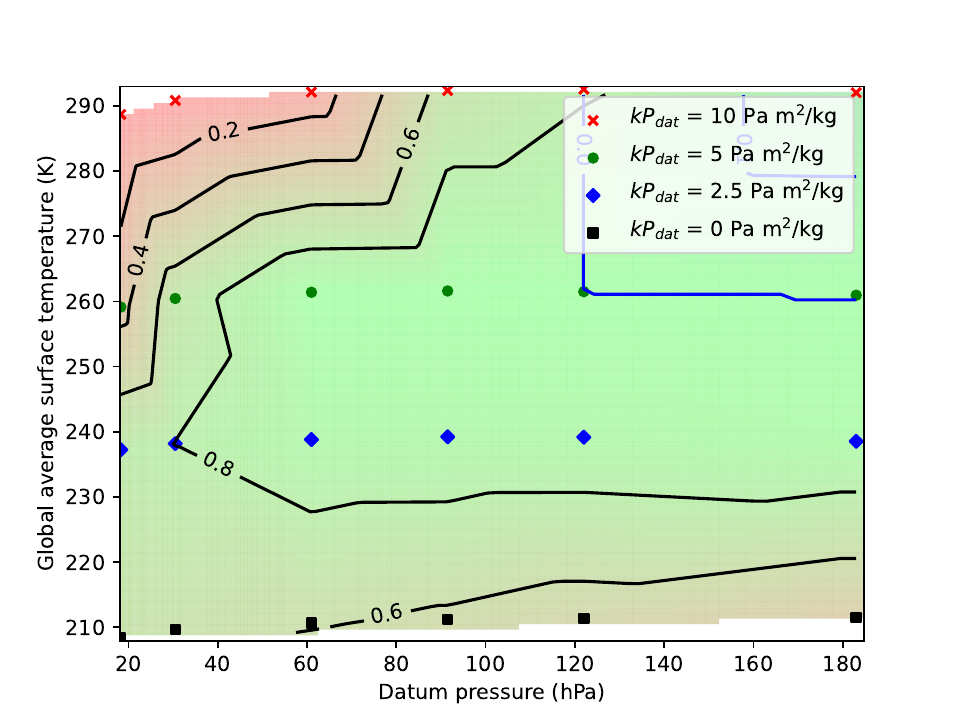}
\caption{Simulated habitability of a warmed Mars as a function of atmospheric pressure (x-axis) and grey gas forcing (symbol colour). Each symbol's position on the y-axis corresponds to the global average temperature for one of our simulations. Black contour lines correspond to the fraction of the surface with temperatures and oxygen levels allowing some~life to~grow; blue contour corresponds to the threshold where, at some places on Mars, humans can live without a~pressure suit. Increasingly green shading shows pressure--temperature conditions that permit photosynthesis but not humans. Increasingly red shading corresponds to lifeless climates.}
\label{temperaturefield}
\end{figure}

\section{Results}\label{sec:Results}

\begin{figure}
\centering
\includegraphics[width=0.8\textwidth,totalheight=0.8\textheight,keepaspectratio]{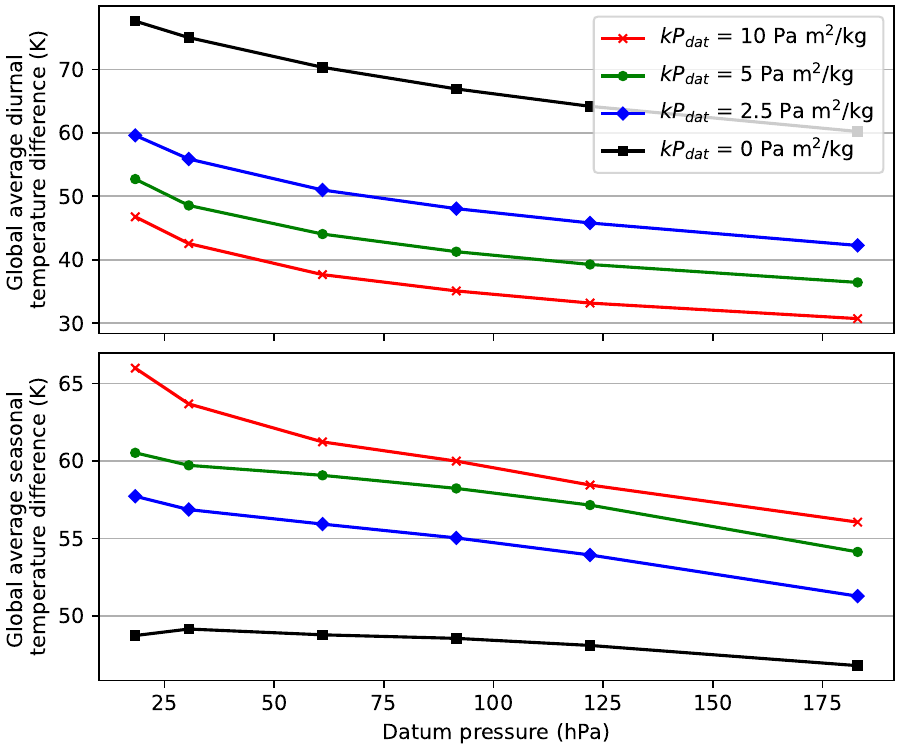}

\caption{As Mars' surface temperature increases with pressure and grey gas forcing, diurnal temperature variations (top panel), defined as the difference between the temperature at 3 pm and the temperature at 3 am local time, decrease due to warmer nights. Seasonal variations (bottom panel) increase with grey gas forcing but~decrease with pressure due to greater heat transfer by advection.}

\label{avcurves1}
\end{figure}

Our grid of models allows us to consider the effects of pressure and the effects of grey gas forcing independently. Within our model, we find \textasciitilde30 K warming when raising $k P_{dat}$ from 0~to~2.5~Pa~m\textsuperscript{2}~kg\textsuperscript{-1}. By contrast, we only find <5~K global warming as pressure increases from $P_{dat}=30.5$~hPa to $P_{dat}=183$~hPa (Fig.~\ref{temperaturefield}); we attribute this warming to the pressure-broadening of the 15 $\mu$m CO\textsubscript{2} absorption band. Warming due to grey gas forcing is strongest at night, when there is no incoming shortwave forcing of the surface. Possible explanations include a) the Stefan--Boltzmann law relating thermal emission to T\textsuperscript{4}, requiring less energy to raise low temperatures than high temperatures, and b)  atmospheric absorption of outgoing longwave radiation hindering radiative cooling of the surface. As grey gas forcing and pressure increase, the elevation dependence of nighttime surface temperatures approaches the atmospheric adiabatic lapse rate (Fig.~\ref{lapserate}). Diurnal temperature ranges decrease by 5 K for every doubling of atmospheric pressure, due mainly to warmer nights (Fig.~\ref{avcurves1}).

\begin{sidewaysfigure}
\includegraphics[width=1\textheight]{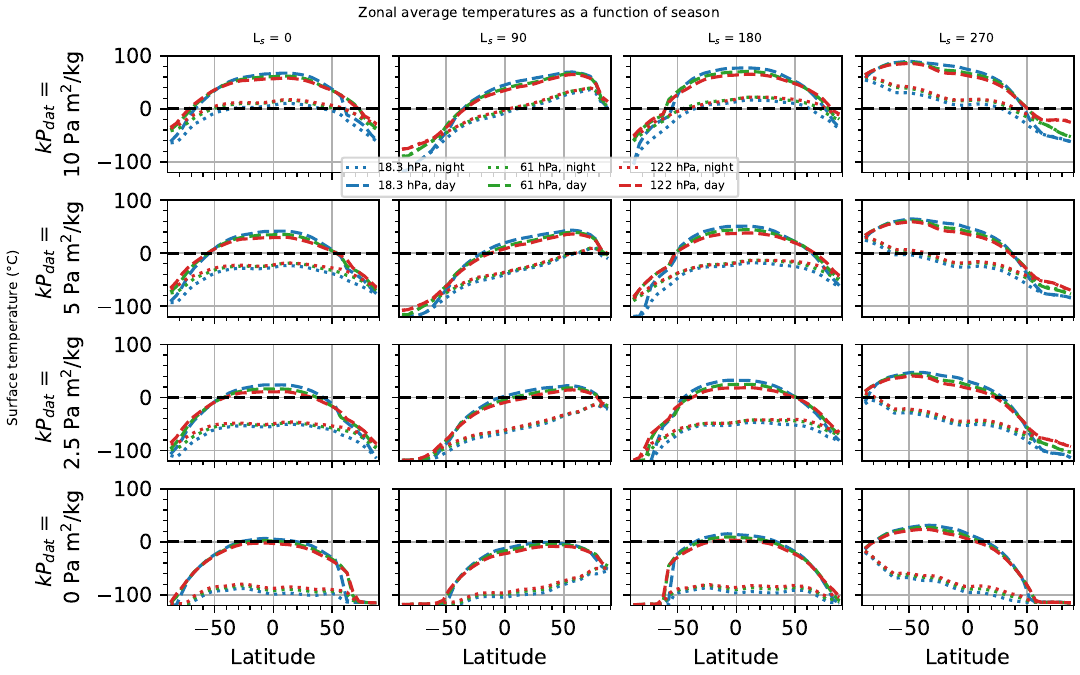}

\caption{Zonal average surface temperatures at four seasons (from left to right: Northern spring equinox, Northern summer solstice, Southern spring equinox, Southern summer solstice), as a function of datum pressure and grey gas forcing. Day and night temperatures
correspond to local times of 3 pm and 3 am respectively. Rising atmospheric pressure warms nighttime temperatures at all latitudes, but cools daytime temperatures
over latitudes where the Sun is directly overhead.}

\label{zonevals}
\end{sidewaysfigure}

\begin{sidewaysfigure}
\includegraphics[width=1\textheight,totalheight=0.7\textwidth,keepaspectratio]{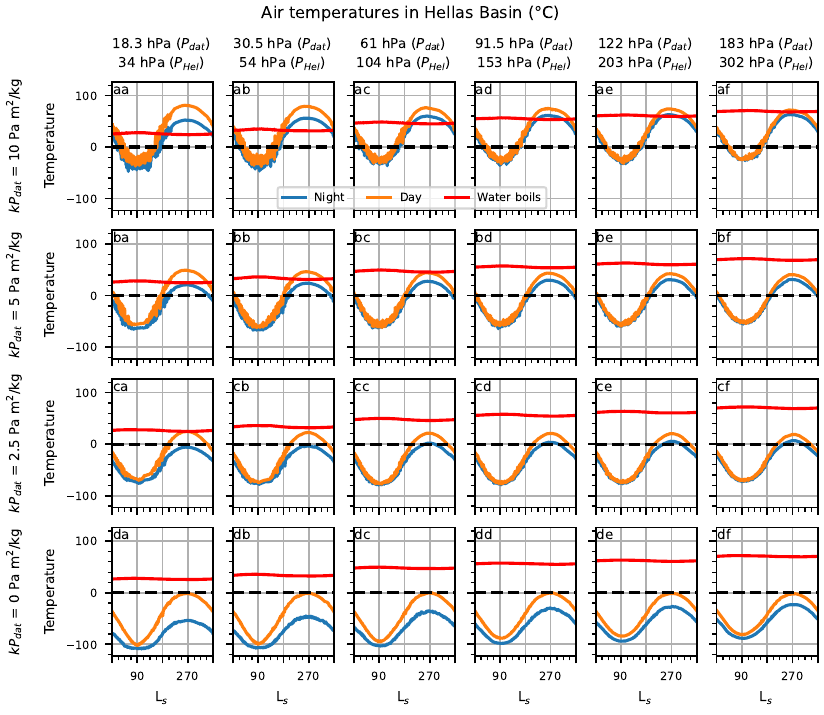}

\caption{Increasing the pressure (left to right panels) and grey gas forcing (top to bottom panels) warms nighttime air temperatures (blue) in Hellas (latitude 42.5° S, longitude 69° E, 6.8 km below datum) near the surface, reducing the day--night temperature range. However, seasonal variations are only slightly reduced. Even at high
atmospheric pressures, high grey gas forcing can cause daytime
highs above the boiling point of water. When~the atmospheric
pressure is 5 times that of present-day Mars, Hellas is at the threshold (Armstrong limit) below which humans would require pressurised suits to prevent blood boiling. Each column header shows the pressure at Mars datum ($P_{dat}$) and that within Hellas ($P_{Hel}$). We plot seasonal variation as a function of solar longitude $L_{s}$ ($L_{s}=90$ is Northern summer solstice, $L_{s}=270$ is Southern summer solstice).}
\label{hellasgraphs}
\end{sidewaysfigure}

 Increased atmospheric pressure damps seasonal temperature variations by lowering annual-peak temperatures (Fig.~\ref{zonevals}). This is likely due to heat redistribution by horizontal advection. In the absence of grey gas warming---which amplifies seasonal temperature variations---the cooler seasonal highs curtail habitability in the mid-latitudes with increasing pressure, and so the area of Mars where organisms can grow shrinks. Nonetheless, even at 183~hPa, the thermal inertia of the atmosphere remains low, and the radiative time constant of the atmosphere---which scales linearly with pressure for a given grey gas forcing \citep{guendelmankaspi2022}---is only $\sim$30~sols, or approximately half that of the Earth's atmosphere. Large seasonal temperature swings in the Southern hemisphere would therefore persist. Life in the Southern mid-latitudes would have to withstand both high diurnal variations in temperature and extreme seasonal variations in temperature due to Mars' eccentric orbit, since perihelion coincides with Southern summer. Increasing pressure would decrease these variations by $\sim$1 K for every 30~hPa increase in atmospheric pressure, but even in the highest-pressure case modelled here ($P_{dat}=183$~hPa, no grey gas forcing), global average seasonal variations are $>$45 K, and 70~K in the Hellas Basin (Fig.~\ref{hellasgraphs}). Seasonal temperature variations in the Northern mid-latitudes would be more moderate, as Mars is further from the Sun during northern summer. On the other hand, the deepest place in the Southern mid-latitudes---Hellas Basin---is the most protected from radiation due to the greater atmospheric column mass \citep{dartnell2007}, and has the largest oxygen partial pressures. Because of the locally higher pO\textsubscript{2}, trees that can tolerate large seasonal temperature swings might first be established in the Hellas Basin. However, the equator and the Northern mid-latitudes---where seasonal variations are~moderate---might be more desirable for humans once the atmosphere becomes breathable.

At current Mars atmospheric pressure, the growth of life on Mars' surface would be limited by radiation and the low (<10 °C) boiling point of water, even at low elevation. This severely restricts the range of temperatures that organisms on the surface would be able to grow in, and hence their spatial range. Photosynthesis would be possible for aquatic cyanobacteria or algae \citep{celekli2024} that can be sheltered from air temperature extremes, for instance in a lake like those in the McMurdo Dry Valleys in Antarctica that~are~covered by translucent ice \citep{kling2020}. 

\begin{sidewaysfigure}
\includegraphics[width=1\textwidth,totalheight=0.8\textheight,keepaspectratio]{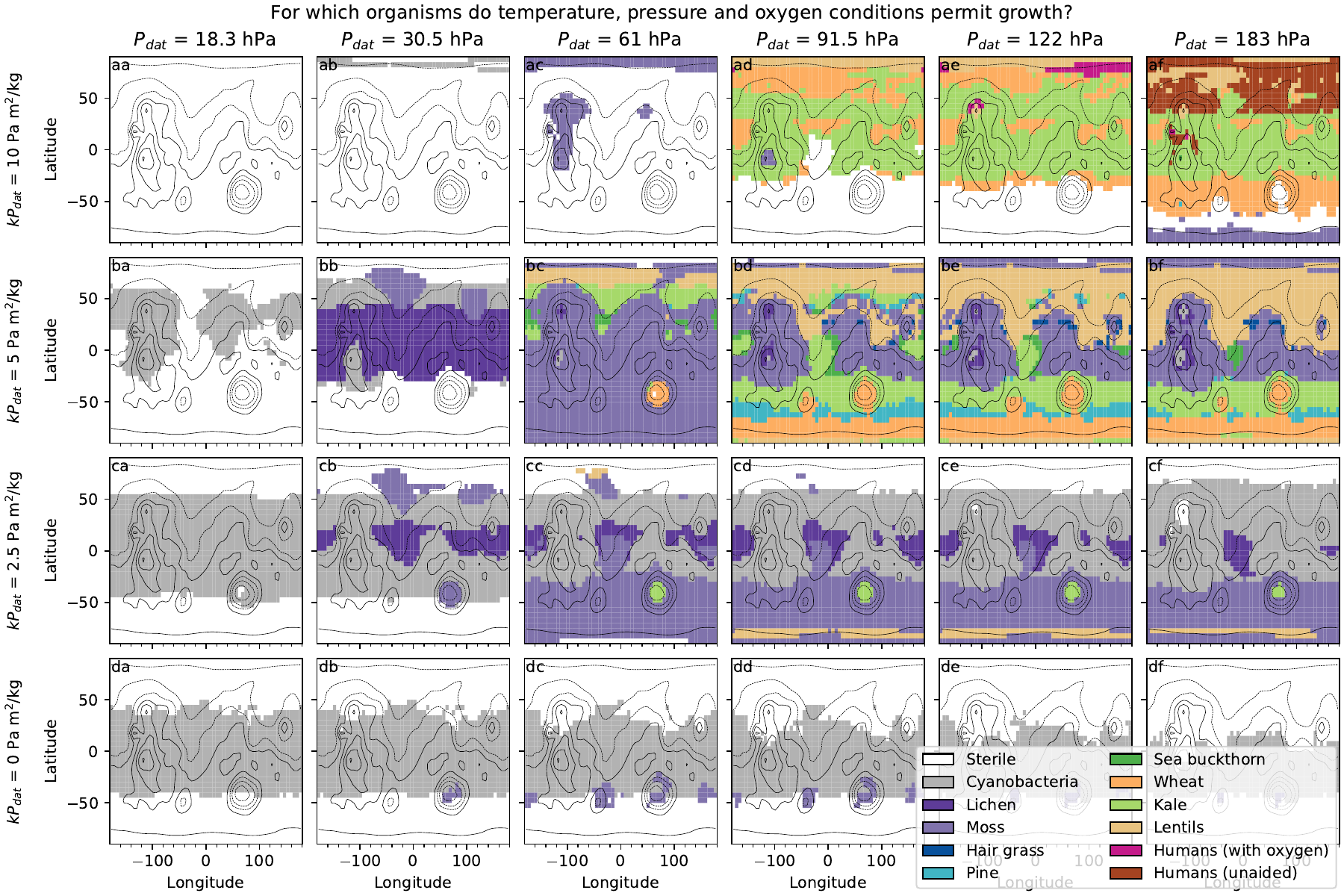}

\caption{Temperature, pressure and oxygen limits on growth of organisms for simulated climates on Mars as a function of atmospheric pressure (left to right) and grey gas forcing (top to bottom). Water-availability and soil-chemistry constraints on life are ignored, so the extents shown are upper limits. At low pressure, the low boiling point of water and low oxygen partial pressure only allow for simple, hardy organisms to grow. As the atmosphere builds up, more complex organisms can be established in the relatively temperate equatorial lowlands. In the legend, organisms are listed from most to least resistant to atmospheric stressors (reading down from the top to the bottom of the~first column and then again from the top to the bottom of the second column), as described in \ref{sec:Minimal-conditions-for} and Tables \ref{organismtable1}-\ref{organismtables2}. Where two or more organisms overlap, we show the constraints for the least resistant one. We distinguish human survival without pressure suits but with oxygen masks (`with oxygen') from human survival when breathing Mars air (`unaided').}

\label{ecopipeline}
\end{sidewaysfigure}

By releasing the entire estimated buried and adsorbed inventory of CO\textsubscript{2}, the atmospheric pressure would approximately triple. This would increase the boiling point of water by $\sim$10 K and greatly increase the area of the planet above the triple point of water (e.g., \citealt{haberle2001}). Taking~into~account only temperature constraints, and assuming optimal grey gas warming, cyanobacteria could live within 50° latitude north and south of the Equator (Fig.~\ref{ecopipeline}) at this stage. Set the amount of grey gas too high, however, and daytime boiling might kill surface organisms---a~grey gas forcing of $k P_{dat}=10$ Pa~m\textsuperscript{2}~kg\textsuperscript{-1} would shrink the habitable area even compared with a scenario with no~artificial~warming. Cyanobacteria would also be constrained by the availability of water. Shallow subsurface water ice is found poleward of 45-50° latitude either side of the Equator \citep{morgan2025}, but not closer to the Equator. This would confine the ecosystem to narrow latitudinal bands, reducing productivity.

\begin{figure}
\includegraphics[width=1\textwidth,totalheight=0.8\textheight,keepaspectratio]{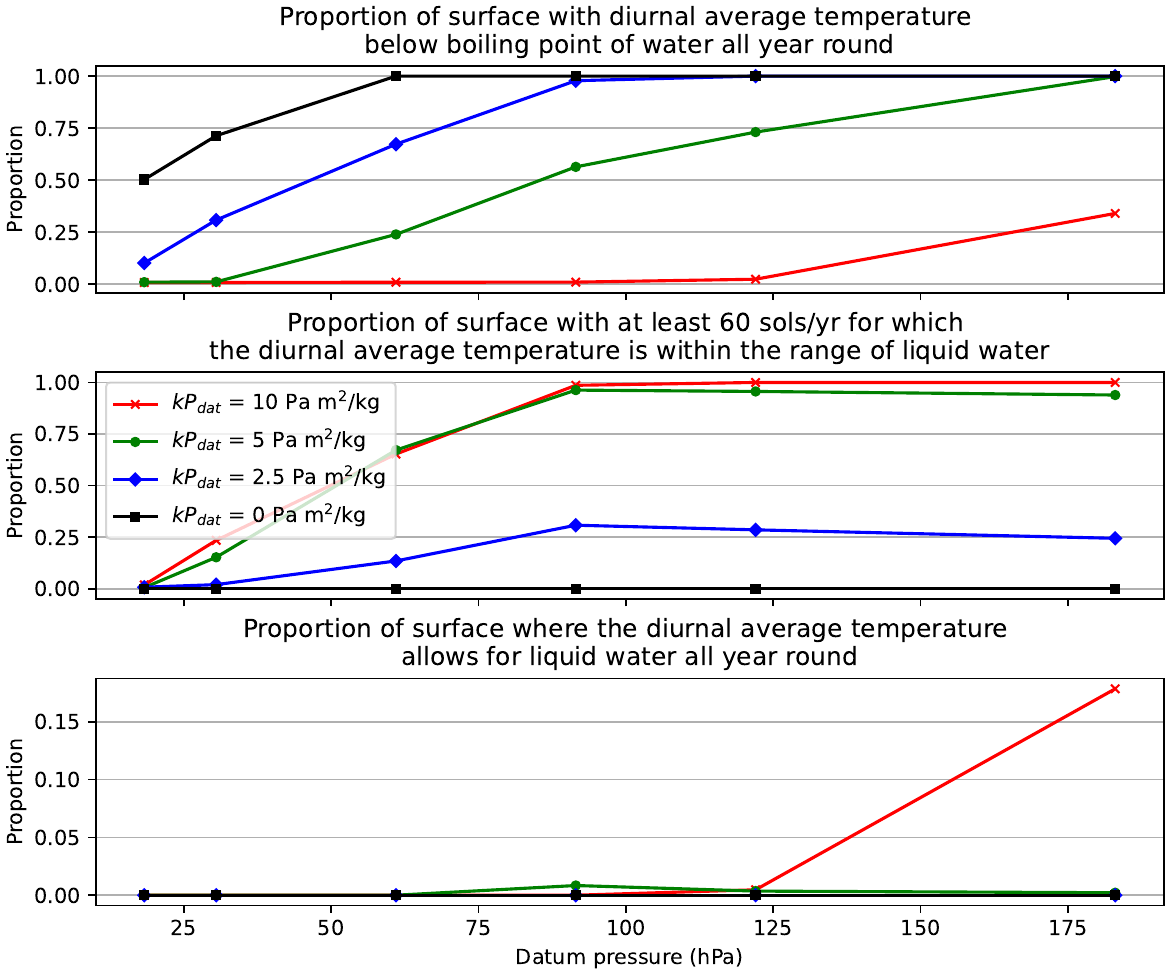}

\caption{As the pressure increases, the boiling point of water rises, allowing for a greater temperature interval with surface liquid water. In turn, this widens the surface area that has seasonal surface temperatures within the~range of liquid water (middle panel), allowing for more opportunities for life to grow on the surface. When~the~atmospheric~pressure is $\geq$15 times that of present-day Mars, small regions of Mars have surface temperatures that permit year-round surface liquid water (bottom panel).}

\label{avcurves2}
\end{figure}

As atmospheric pressure increases, water is less likely to~boil during the day and freeze at night. Nonetheless, even at atmospheric pressures 5 times that~of~present-day Mars (i.e., at 30~hPa), at most a quarter of the Martian surface has $>$60 sols per Mars year with diurnally-averaged temperatures above 0 °C without
exceeding the boiling point at a local time of 3 pm (Fig.~\ref{avcurves2}). This would, however, allow for some lichens and, when the oxygen partial pressure permits, mosses, to become established in the Hellas Basin, and then at $P_{dat}\sim60$
hPa, expand across the Southern mid-latitudes and then almost the whole planet. These organisms could only grow during short parts of the day between daily cycles of freezing and desiccation through boiling.

As the surface pressure approaches 10 times that of present-day Mars, the oxygen content would be consistent with crops, flowers, and forests, although hydrological or soil-chemistry constraints (not included in our model) might prevent this in practice. Hellas would become the first habitat for trees such as \emph{Pinus sibirica} (Fig.~\ref{ecopipeline}), assuming optimal grey gas forcing \citep{wendland2026}. Automated agriculture may be possible in the Southern mid-latitudes for part of the year (provided crops are resistant to the UV environment). At this point, humans would not need pressure suits in lowland regions of Mars near the Equator, although ultraviolet and galactic cosmic rays would still be hazardous (\ref{sec:Minimal-conditions-for}) and oxygen masks would still be needed. At $P_{dat}\sim100$~hPa (\textasciitilde 15~times the current Mars pressure), small low-lying regions near the Equator would have diurnal average temperatures above freezing year-round. By~20--30 times present-day Mars pressure, large parts of Mars would have human-breathable air, with $>$15\% of the Martian surface having the temperature and pressure conditions for year-round liquid water.

\begin{figure}
\centering
\includegraphics[width=0.75\textwidth]{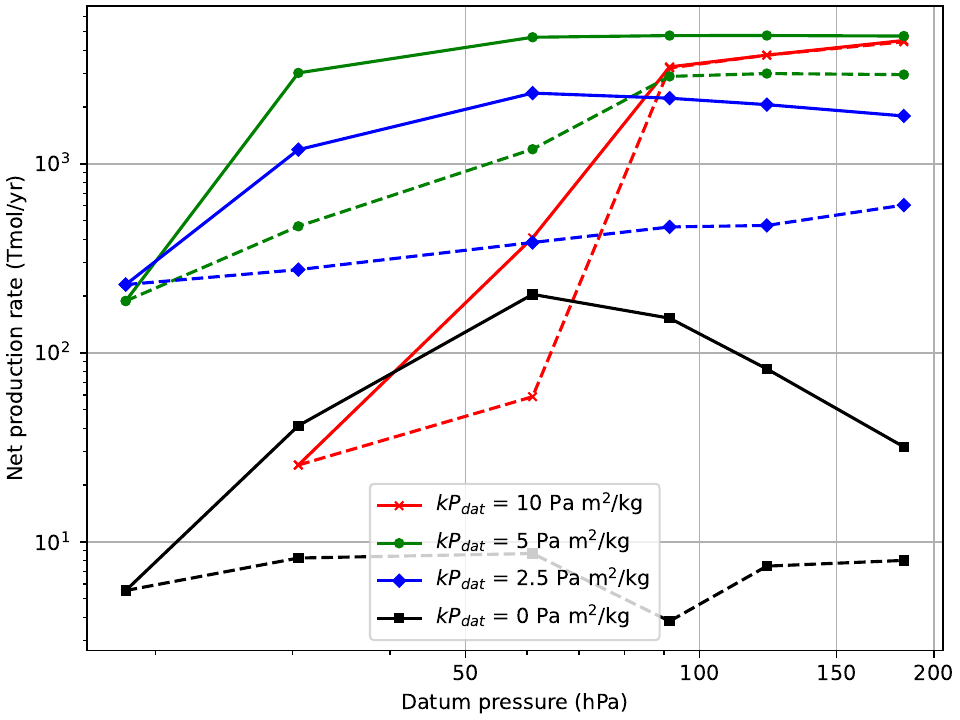}
\caption{Estimated net production rate of oxygen over the area that would sustain organisms (Fig.~\ref{ecopipeline}) if~the~production rate is assumed to be 33 mol/m\protect\textsuperscript{2}/yr for plants and lichens (equivalent to tussock grass, \citealt{shaverchapin1991}) and 26 mol/m\protect\textsuperscript{2}/yr for cyanobacteria  (equivalent to 50\% of industrial production rates, \citealt{Verseux2016}). Dashed lines correspond to the assumption that the productivity of regions in which only mosses and lichens can grow is 0.8 mol/m\protect\textsuperscript{2}/yr, equivalent to the lower productivity end of tundra environments on Earth \citep{whittaker1975communities}. Cyanobacteria are assumed to only grow poleward of 45° latitude, where data suggest near-surface ice. The local dip in the productivity when $kP_{dat} = 0$ and $P_{dat} = 91.5$~hPa is due to the spatial resolution of the model; at these pressures, only a few pixels poleward of 45° permit cyanobacteria.}

\label{oxygenproduction}
\end{figure}

We do not know how long oxygenation would take. The harsh soil environment is likely to slow net oxygen production, especially in early stages of terraforming. However, if we assume (1) net primary productivity of 400~g/m\textsuperscript{2}/yr of total carbon photosynthesised within the area that we estimate could sustain plants and lichen (equivalent to tussock grass ecosystems on Earth: \citealt{shaverchapin1991}), and~(2)~312~g/m\textsuperscript{2}/yr within the area that could only sustain cyanobacteria (equivalent to 50\% of industrial cyanobacteria production rates in \citet{Verseux2016}, and where we further assume, for the purposes of this O\textsubscript{2} production calculation, that cyanobacteria are restricted to latitudes poleward of 45° where subsurface ice is inferred by \citealt{morgan2025}), this would translate to at most $7\times10^{12}$ kg/yr of O\textsubscript{2} produced at 18.3~hPa pressure, rising to at most $10^{14}$ kg/yr of O\textsubscript{2} at 30.5~hPa (Fig.~\ref{oxygenproduction}). Thus, between 500 and 7000 Earth years would be required to increase the atmosphere from 3 times to 5 times current Mars pressure, assuming atmospheric CO\textsubscript{2} is~completely recycled during the O\textsubscript{2} production process (e.g., electrons transferred via iron reduction or sulphate reduction). In reality, the availability of water, nutrients, UV irradiation and hypoxia would reduce oxygen productivity. As atmospheric CO\textsubscript{2} is depleted, photosynthesis would also be limited by the need to recycle sequestered carbon back into the atmosphere using iron and sulphate reduction (\ref{sec:How-much-oxygen}). Building up an~oxygen atmosphere would therefore take several thousand years or even more, unless accelerated using synthetic biology, cell-free methods \citep{hann2022}, or abiotic water-splitting.

\section{Discussion}\label{sec:Conclusions-and-future}

Our work contains several assumptions and uncertainties. We assume irrigation, wells, rainfall, streams, or groundwater flow supply water
in regions that currently lack shallow-subsurface ice (within approximately 45° latitude north and south of the Equator, 
\citealt{morgan2025}). Surface vegetation would need more soil-water than on Earth,
because the low pressure of the Martian atmosphere speeds up evapotranspiration losses \citep{hecht2002}. In turn, evapotranspiration would modify Mars' water
cycle. Clouds would alter the planet's radiative balance and moderate diurnal (and perhaps seasonal) temperature swings \citep{braude2026arxiv}. The behaviour of the dust cycle and
cloud microphysics on a warmed and oxygenated Mars merits more research and model intercomparisons, for example because dust could cool the surface and block out sunlight needed for photosynthesis.

We also assume that atmospheric build-up is from O\textsubscript{2}, not N\textsubscript{2}. Low partial pressure of N\textsubscript{2} ($\sim$20~Pa) could make nitrogen fixation sluggish \citep{klingler1989}, a potentially serious problem for a Mars biosphere. Closing the terraformed Mars nitrogen cycle is essential to sustain a biosphere, and would require either abiotic nitrogen fixation (which underpins 1/2 of Earth's agriculture) or biological nitrogen fixation (which underpins the other 1/2 of Earth's agriculture). Extrapolating from the only field site with data \citep{Stern2015}, only $\sim$1 Pa of N\textsubscript{2} could be released from nitrates in the soil. However, some models \citep{manning2008} predict that a~subsurface nitrate reservoir sequesters 60~hPa N\textsubscript{2}. There are no known sources of other inert gases on Mars that would substantially increase Mars' atmospheric pressure. 

An artificially warm Mars with a higher-O\textsubscript{2} atmosphere would
differ from Earth. Seasonal variations in temperature would remain
extreme due to the combined obliquity and eccentricity forcing, and
be drawn out over a year that is almost twice as long as an Earth
year. Water would remain a precious resource, as Mars' known water inventory amounts to a 30 m global equivalent layer \citep{jakoskyhallis2024}, much less than for Earth. This scarcity could be exacerbated by a potential `icy highlands' effect, where much of Mars' water accumulates in the cold highlands and becomes difficult to melt. While we do not predict this to occur on a warm Mars at present-day atmospheric pressures \citep{braude2026arxiv}, it could become significant at higher atmospheric pressures and grey gas forcings where the vertical temperature profile approximates a dry adiabat (Fig. \ref{lapserate}, \citealt{fan2023}). A terraformed Mars
would not be a second Earth, but a different
planet that opens new scientific questions about the breadth of stable
climates that can sustain habitable worlds (e.g., \citealt{Wolf2017}), the search for extraterrestrial intelligence
(e.g., \citealt{haqqmisra2022,vidal2026}), the stability of biospheres
\citep{nelson2021,Chen2024}, and
the adaptability of organisms to environmental stressors alien to
those found on the Earth \citep{coleine2025}.

\section{Conclusions}\label{sec:Conclusions}

We used the MarsWRF 3-D Global Climate Model to model a warm and oxygenated Mars. We considered CO\textsubscript{2}--O\textsubscript{2} atmospheres, from an average surface pressure of 18.3~hPa and 97\% CO\textsubscript{2} (equivalent to Mars following outgassing of all adsorbed
and buried CO\textsubscript{2}) to 183~hPa (a 90\% O\textsubscript{2}--10\% CO\textsubscript{2} atmosphere with near Earth-like oxygen partial pressures). Increased atmospheric pressure and grey gas warming moderate Mars' extreme diurnal temperature variations by contributing more warming in the night than the day. However,  seasonal temperature variations, which are amplified by Mars' high eccentricity \citep{laskar2004}, would stay large. Our model predicts the Hellas Basin to be the most favorable for life in earlier stages of Mars' atmospheric evolution. Here, seasonal warming and shielding from radiation are greatest, and the atmospheric pressure widens the temperature range over which water can remain liquid. Microbes could be established once the surface pressure is only a few times that of present-day Mars. Climates permitting more complex organisms that are more sensitive to temperature extremes would be possible only later in Mars' atmospheric evolution, with a preference for lowlands near the Equator and the Northern tropics (where seasonal temperature variations are more modest than in the Southern hemisphere). Elsewhere, growing seasons would be short. Neglecting UV and GCR constraints (\ref{sec:Minimal-conditions-for}),
we predict that crops could grow once Mars has reached an average
surface pressure 10--15~times that of its present-day
value (\textasciitilde 60~-~100~hPa), which gives enough oxygen
for plants in lowland areas of Mars to grow and reproduce.
Above 150~hPa datum pressure, humans would be able to breathe outdoors at the lowest elevations.

Mars can be warmed, but, assuming only the use of resources on Mars, we do not yet know if~it~is~practical for Mars to be oxygenated. The main open questions include: 
\begin{itemize}
  \item Can significantly more moles of oxygen be added to Mars' atmosphere than moles of CO\textsubscript{2} that are outgassed?
  \item Can life adjust to the salty soil chemistry and freeze--thaw cycles of a warmed Mars, through synthetic biology or bioremediation?
  \item Would surface UV levels make photosynthetic oxygen build-up very slow on civilisation-relevant timescales? 
\end{itemize}

\noindent If Mars is oxygenated, core attributes of the climate/habitability would be:
\begin{itemize}
  \item Extreme seasonal temperature variations in the Southern mid-latitudes that would continue to pose challenges for a complex biosphere.
  \item More moderate temperature variations in the Equatorial and Northern Lowlands, perhaps more suitable for human habitation assuming sufficient warming.
  \item A climate that would likely still have much less water coverage than the Earth.
\end{itemize}

\begin{acknowledgments}
Particular thanks to Mark Richardson, whose support in the application of MarsWRF to this project was essential to the success of the work. This work used resources
from the University of Chicago's Research Computing Center, and~the~NASA High-End Computing (HEC) Program through the NASA Advanced Supercomputing
(NAS) Division at Ames Research Center. A portion of this work was
conducted at the~Jet~Propulsion~Laboratory, California Institute of
Technology, under contract with NASA. We~thank Erika DeBenedictis,
Devon Stork, Sukrit Ranjan, Ben Tutolo, and the PlanetWRF development team for their support.
This work was partly funded by Astera Institute. 

\bibliography{mars}
\end{acknowledgments}

\appendix

\subsection*{Open access statement: }

All model output is available on Zenodo (doi:10.5281/zenodo.23306237), including input files that allow for reproducibility. The MarsWRF
source code can be made available by Aeolis Research pending scientific
review and a completed Rules of the Road agreement. Requests for the
MarsWRF source code should be submitted to mir@aeolisresearch.com.

\pagebreak{}

\include{end_state_arxiv_supplementary_draft3}
\end{document}

%% file: end_state_arxiv_supplementary_draft3.tex
\title{Supporting information for `The climate and potential habitability of a warm and oxygenated Mars'}

\authors{Ashwin S. Braude, \affil{\selectlanguage{english}%
1}\\
Edwin S. Kite, \affil{\selectlanguage{english}%
1,2}\\
Preston Cosslett Kemeny, \affil{\selectlanguage{english}%
3}\\
Raymond T. Pierrehumbert, \affil{\selectlanguage{english}%
4,5}\\
Alexandre Kling, \affil{\selectlanguage{english}%
1}\\
and Michael A. Mischna, \affil{\selectlanguage{english}%
6}}

\affiliation{1}{Astera Institute, Emeryville, CA, USA.}

\affiliation{2}{University of Chicago, Chicago,
IL, USA (kite@uchicago.edu).}

\affiliation{3}{Woods Hole Oceanographic Institution, Woods Hole, MA, USA.}

\affiliation{4}{MIT, Cambridge, MA, USA.}

\affiliation{5}{University of Oxford, Oxford, UK.}

\affiliation{6}{Jet Propulsion Laboratory, California Institute of Technology, Pasadena,
CA, USA.}

\setcounter{figure}{0} 
\renewcommand{\thefigure}{S\arabic{figure}}
\renewcommand{\thetable}{S\arabic{table}}\pagebreak{}

\section{Defining minimum conditions for habitability}\label{sec:Minimal-conditions-for}

A biosphere requires life to grow. Many life-forms can survive extreme conditions by shutting down their metabolism (e.g., \citealt{tirumalai2025}), and resuming
it once better conditions return. For example, the desert moss \emph{Syntrichia caninervis} can grow after immersion in liquid nitrogen \citep{Li2024} at -196°C. However, organisms cannot contribute to Mars ecosynthesis unless they can grow and reproduce, and no known unicellular organism can grow below -20 °C \citep{clarke2014}. For this reason, we distinguish two separate temperature constraints for each organism: first, $T_{s}$, defined as the temperature extremes the organism can survive for at least one day, but not necessarily be active in, and second, a narrower temperature range $T_{L}$, corresponding to an extended period of time (assumed to be $\geq$60 sols, unless otherwise specified) within which the organism can grow and metabolise.

Because of the high seasonal and diurnal swings in temperature, extreme heat may be as concerning for life during Mars terraforming as extreme cold, especially in the Southern mid-latitudes where seasonal temperature swings are greatest. Plants alleviate heat stress through transpiration \citep{hasanuzzaman2013,kumar2025}, but this risks dehydration, especially on~Mars where evaporation is faster due to the low atmospheric pressure \citep{hecht2002}. Unless otherwise specified, we therefore assume an upper temperature bound on survival equal to the boiling point of water; in practice, survival and growth in high temperatures would depend on water availability. While near-surface ice is~currently found poleward of 45° latitude on either side of the equator \citep{morgan2025}, active hydrology could spread water more broadly \citep{gauvain2026, braude2026arxiv}. Because not much work has been done on water availability on a~terraformed Mars, we neglect this effect on habitability in this work.

Mars has additional stressors that would severely inhibit
growth. One is~low oxygen partial pressure (\emph{hypoxia}). While many microbes can thrive without oxygen, and some lichens and fungi tolerate hypoxia \citep{devera2014},
plants need between
20 and 60 hPa to allow for~germination and root respiration (reviewed
in \citealt{graham2004}). Given sufficient O\textsubscript{2}, plants can tolerate low total atmospheric pressure (hypobaria) \citep{he2007}. CO\textsubscript{2} on Mars would also not be a limiting factor in plant growth: present-day CO\textsubscript{2} partial pressures on Mars at datum are approximately twenty times that of the pre-industrial Earth at sea level.

\begin{figure}
\includegraphics[width=1\textwidth,totalheight=0.9\textheight,keepaspectratio]{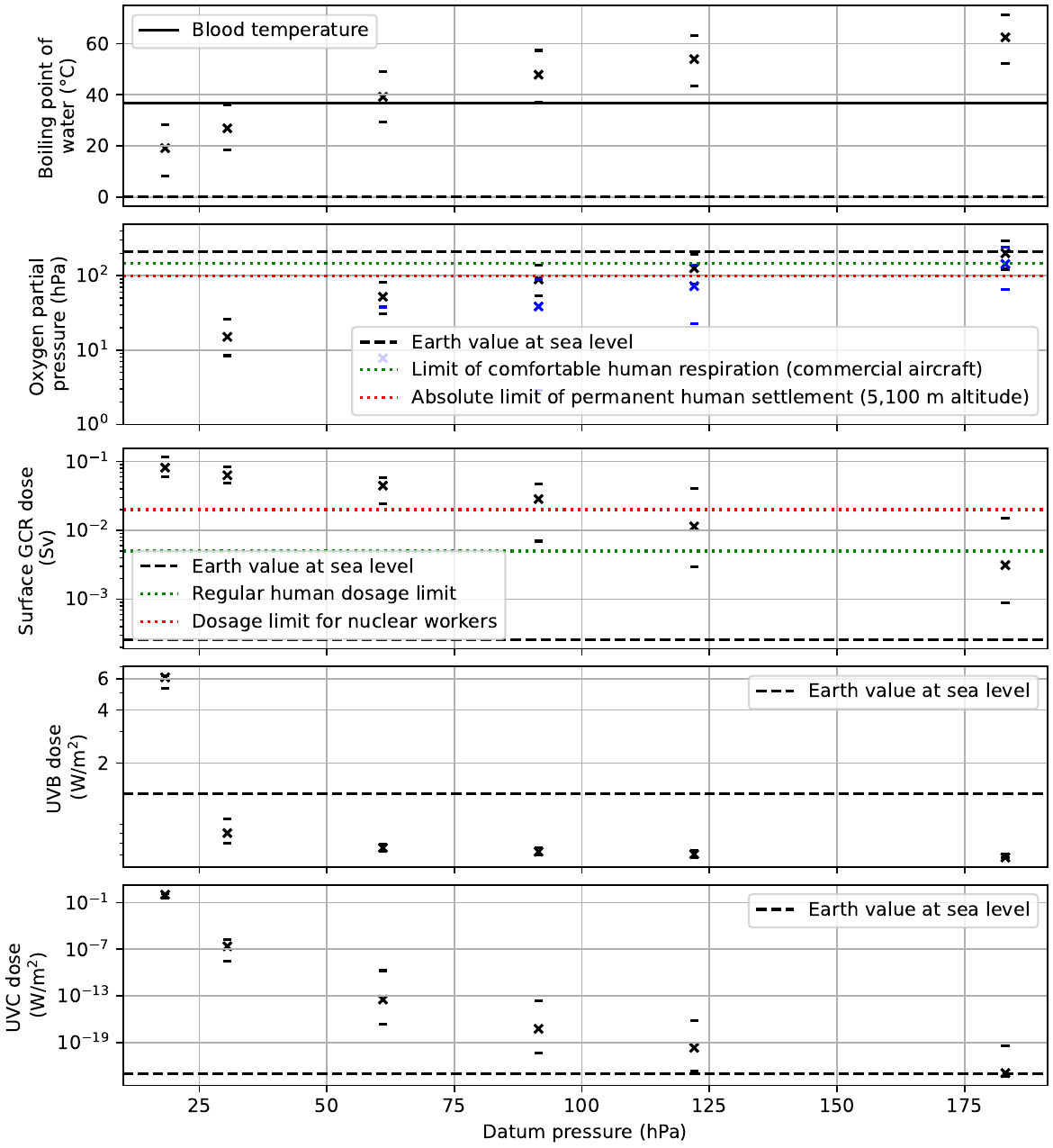}

\caption{Constraints on life. In each panel, the black crosses show the value
at datum elevation for each scenario, and the short
horizontal black lines correspond to the lowest altitude in Hellas (-6.8 km, most habitable for these parameters) and the highest altitude in
Tharsis (+8.2 km, least habitable for these parameters). A~constant 18 hPa
partial pressure of CO\protect\textsubscript{2} is~assumed. (Top panel) Boiling point of water, with the temperature of human blood shown for reference (horizontal
black line). (Second panel) Oxygen partial pressure and (blue crosses) equivalent inspired oxygen partial pressures calculated according to the Equivalent Air Altitude Model, discussed in (e.g.) \citet{conkinwessell2008}. Also shown: minimum inspired oxygen partial pressure for human survival (red dotted line) and minimum inspired oxygen partial pressure for comfortable
breathing (green dotted line). 
(Third panel) Galactic Cosmic Ray yearly daytime equivalent surface
dose rate modelled according to \citet{dartnell2007}, with approximate
background limits of 5 mSv shown in green, and the GCR dosage limit
for nuclear workers (20 mSv per year, as~recommended by \citealt{icrp2007})
shown in red. Black dashed line shows the approximate value at Earth
sea~level. (Second-from-bottom panel) UVB daytime surface irradiation
and (bottom panel) UVC daytime surface irradiation modelled according
to \citet{segura2003}, as a function of average surface pressure,
with black horizontal dashed line corresponding to the value at Earth
sea~level. }
\label{uvgcrgraphs}
\end{figure}

\begin{sidewaysfigure}
\includegraphics[width=1\textwidth,totalheight=0.8\textheight,keepaspectratio]{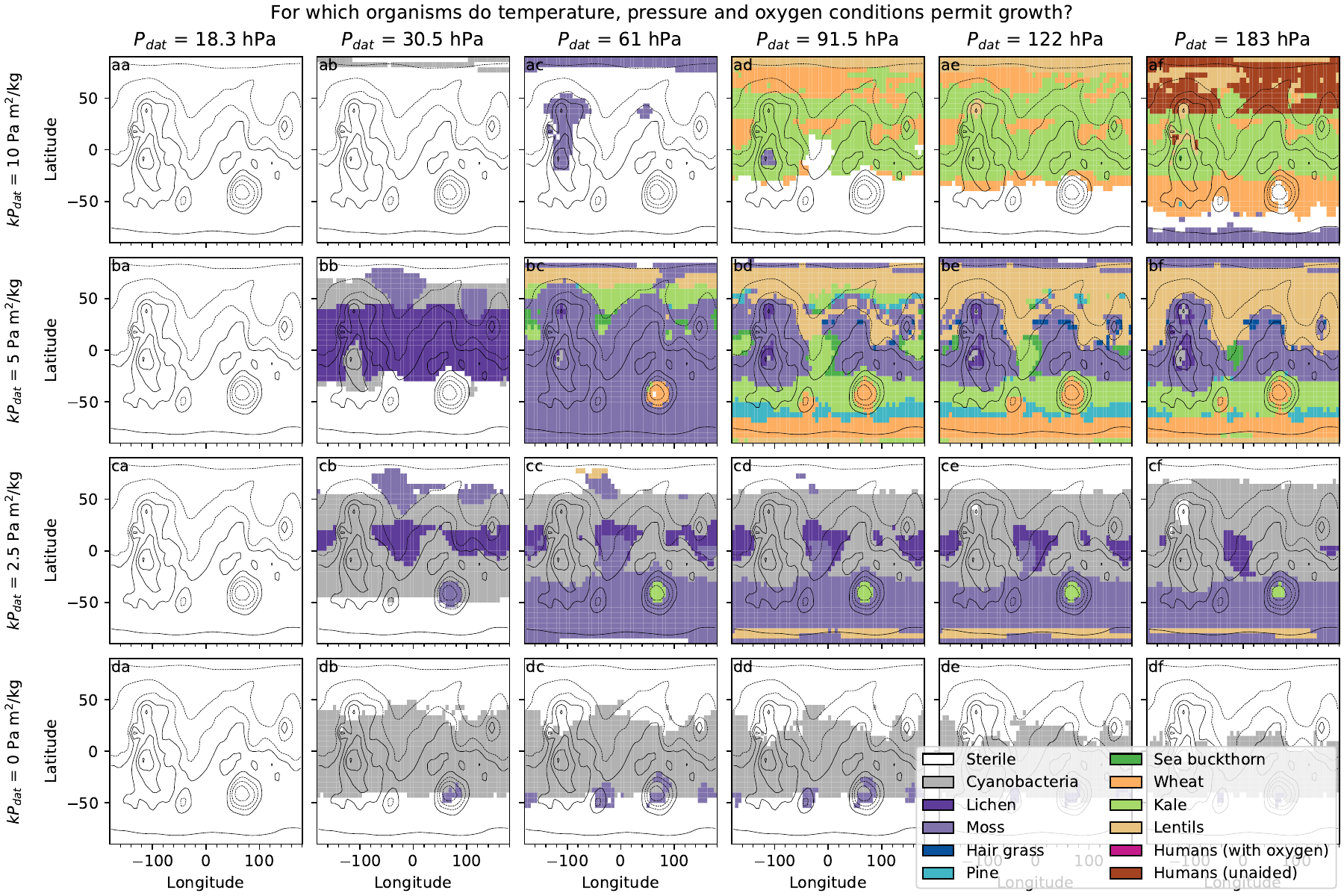}

\caption{As Fig. \ref{ecopipeline}, but with an additional maximum constraint
of 5 mSv of galactic cosmic radiation for~human survivability and
0.0025 W m\protect\textsuperscript{-2} of UVC radiation for the survivability
of cyanobacteria. These constraints limit bacterial growth
to areas protected from radiation (e.g., in sheltered outcrops or in
lakes covered by ice) when $P_{dat}<30$ hPa, and human settlement
to $P_{dat}>180$ hPa. With these assumptions, Galactic Cosmic Rays (GCR) are~a more limiting constraint on human settlement than oxygen partial pressure.}

\label{ecopipeline-1}
\end{sidewaysfigure}

Life on the
surface of Mars would also be subjected to UV radiation, Solar Energetic Particles (SEPs), and Galactic
Cosmic Rays (GCRs) (Fig. \ref{uvgcrgraphs}). We neglect radiation constraints
on habitability in the main text, but assess two effects of radiation on life in Figs.
\ref{ecopipeline-1}-\ref{ecotables-1}: 1) the effect of UVC radiation
on the viability of cyanobacteria during early Mars terraforming and
2) human tolerance to GCR. 

UVC radiation (100--280 nm) is~blocked by Earth's ozone layer, but enough  reaches the surface of Mars (1.5--2 W m\textsuperscript{-2}) to kill unshielded microbes living
there (e.g., \citealt{ViudezMoreiras2025, vaishampayan2012}). On a terraformed Mars, absorption of UV radiation could be from engineered aerosols (e.g., \citealt{richardson2025}), or a biologically produced surface screen, or from O\textsubscript{3} that forms in the~atmosphere as O\textsubscript{2} builds up.  We relate the surface flux of UVC to oxygen partial pressure of model atmospheres for Earth-like planets around Sun-like stars from
Table 2 of \citet{segura2003} adjusted for Mars' lower insolation,
neglecting uncertainties in the combined effects of the water and
chlorine cycles which could dominate ozone destruction \citep{Metz2005}.
In Figs. \ref{ecopipeline-1}-\ref{ecotables-1}, we impose a UVC
tolerance upper limit of 0.0025 W m\textsuperscript{-2} for cyanobacteria.
We base this limit on the work of \citet{baque2013}, which showed
that cyanobacteria remained viable after 70 J m\textsuperscript{-2} of UVC irradiation 
over 24 hours. We have converted this figure to an
irradiation rate assuming 12 hours of daylight, and scaling by
a factor of $\pi/2$ to account for variation in insolation throughout
the~day. Constraints on UVC tolerance are from \citet{sanchez2014} for lichens and \citet{Li2024} for mosses. We neglect
UVB radiation (280 - 315 nm) constraints on survivability, even though
they would likely be limiting at stages in Mars'
atmospheric evolution when there is~sufficient ozone to~screen~out harmful levels of UVC radiation, but not to screen out harmful levels of UVB radiation. 

In Figs. \ref{ecopipeline-1} and \ref{ecotables-1}, we use a GCR limit for humans on~Mars of 5 mSv/year. This is~more than the 1 mSv recommended maximum yearly exposure from artificial sources for the general public \citep{icrp2007}, but this supplements natural radiation levels on Earth of a few mSv/yr. GCR attenuation by Mars air is~taken from \citet{dartnell2007}, and doses are multiplied by 2 when~converting to equivalent dose rates
to account for the high proportion of proton radiation \citep{icrp2007}.

We neglect other aspects of Mars' soil (reviewed in
e.g., \citealt{schuergernicholson2006,berry2010effects}) that could
prevent the growth of life, including salinity and the presence of perchlorates. We~assume~soil is~remediated through
synthetic biology, for instance perchlorate-reducing bacteria
(see \ref{sec:How-much-oxygen}). Although many extremophiles can survive exposure to individual stressors of~the~Mars environment, the Martian
environment would subject organisms to multiple simultaneous stressors
(e.g., \citealt{CaroAstorga2024}), making synthetic biology likely necessary (although we do not know if it would be sufficient).

\section{Harnessing biogeochemistry to oxygenate Mars}\label{sec:How-much-oxygen}

Mars has a $>$30 m Global Equivalent Layer (GEL) of water ice \citep{jakoskyhallis2024}, enough for conversion to $\sim$1 bar of atmospheric oxygen. Such a conversion, however, requires transferring electrons from water to another element or molecule (eq. \ref{eq:scheme1}) that is~then removed from interaction with the product oxygen. Regardless of whether atmospheric oxygenation is~done abiotically or (discussed below) biologically, a key question is~the fate of the electron sink. In this section, we present possible pathways to atmospheric oxygen build-up through schematic geochemical reactions (as opposed to specific biochemical reaction steps). For each estimate of oxygen partial pressure, we assume a conversion ratio of 3.54 $\times$ 10\textsuperscript{-14} Pa of oxygen partial pressure per kg of O\textsubscript{2} released into an atmosphere of 44 g mol\textsuperscript{-1} mean molecular weight. As the atmosphere becomes more oxygen-dominated (i.e. of 32 g mol\textsuperscript{-1} mean molecular weight), this value tends to 2.58 $\times$ 10\textsuperscript{-14} Pa of oxygen partial pressure per kg of O\textsubscript{2}. The estimated partial pressures in this appendix should therefore be reduced by up to 25\% if emitted later in Mars' oxygenation process.

The abiotic electrolysis and photolysis of water produce both O\textsubscript{2} and H\textsubscript{2} (eq. \ref{eq:scheme2}). Abiotic water-splitting is~more energy-efficient than photosynthesis \citep{blankenship2011}; but is difficult to scale as solar panels cannot self-replicate. The two chemical products must then be separated to~prevent back-reaction; this could occur through the capture of H\textsubscript{2} and either active release above the homopause (from which escape to space is~assured), or underground sequestration. Either would result in a net release of O\textsubscript{2} into the Martian atmosphere. Indeed, although it is~now thought that high levels of oxygen in Earth’s atmosphere reflect oxygenic photosynthesis, 20th-century scientists hypothesised that, at~times, atmospheric O\textsubscript{2} rose due to abiotic water photolysis \citep{brinkmann1969,lovelocklodge1972}. 

Turning to biology, oxygenic photosynthesis splits water by harnessing light energy to fix CO\textsubscript{2} into organic matter (approximated here as CH\textsubscript{2}O; eq. \ref{eq:scheme3}). On~Mars, insufficient CO\textsubscript{2} exists to support the production of global O\textsubscript{2} levels required for humans to~breathe. \citet{jakoskyedwards2018} suggest there is~only enough known CO\textsubscript{2} for \textasciitilde20 hPa of O\textsubscript{2}, which would restrict a global biosphere to lichens and simple plants. \citet{broquet2020} and \citet{buhlerpiqueux2021} suggest indirect evidence for additional CO\textsubscript{2} reservoirs totalling \textasciitilde100 hPa, barely sufficient at lower elevations for humans to~breathe comfortably. Moreover, Mars’ known carbonate reservoirs are small compared to the Earth's \citep{edwards2015,kite2025}, while volcanic outgassing (which would include mantle CO\textsubscript{2}) is~undetectably slow on present-day Mars \citep{braude2022so2}. Thus, oxygenic photosynthesis is~unlikely to~produce the desired quantities of O\textsubscript{2} directly. However, the organic matter produced through oxygenic photosynthesis might be harnessed to facilitate a net transfer of electrons from water to other compounds. That is, by coupling oxygenic photosynthesis with anaerobic respiration, oxygen partial pressures $>$20 hPa might be reached. We investigate the possibility that perchlorate (eq. \ref{eq:scheme4}), ferrihydrite (eq. \ref{eq:scheme6}), and ferric sulphate (eq. \ref{eq:scheme10}) present on the modern Martian surface might all be harnessed to oxidise organic matter and, in net, release oxygen into the atmosphere (eqs. \ref{eq:scheme5}, \ref{eq:scheme7}, and \ref{eq:scheme11}). Other~combinations of biogeochemical reactions can also result in net oxygen production \citep{kemeny2024}.

The Martian soil has 0.4-0.6 wt\% 
\citep{hecht2009} perchlorates at the one site that has been measured. Perchlorates are~toxic for many forms of life. Bioremediation by adding perchlorate-reducing bacteria (eq. \ref{eq:scheme4}; e.g., \citealt{coatesachenbach2004}) to the soil might allow for a locally more complex ecosystem while recycling CO\textsubscript{2} that could then be photosynthesised again. Assuming a globally-averaged soil thickness of 3 m \citep{2023LPICo2806.1843R} with a bulk density of 1500~kg~m\textsuperscript{-3} \citep{piqueuxchristensen2011}, metabolising the 0.6 wt\%
of ClO\textsubscript{4}\textsuperscript{-} in the soil would release at most 2.5 $\times$ 10\textsuperscript{15} kg (\textasciitilde 90 Pa) of O\textsubscript{2}, insufficient for  plant or human life.

\citet{valantinas2025} proposed that the main iron oxide in Mars dust is~ferrihydrite (idealised here as Fe\textsubscript{5}O\textsubscript{8}H). Ferric iron is~an electron acceptor, and the reduction of ferrihydrite might contribute to the recycling of CO\textsubscript{2}, allowing more photosynthetic production of oxygen. Assuming a global dust inventory equivalent to a 3 m global average layer \citep{ojha2018} of density 2600~kg~m\textsuperscript{-3} \citep{lopezcayuela2024}, with $\geq$10 wt\%
ferrihydrite \citep{valantinas2025}, complete reduction of Fe\textsubscript{5}O\textsubscript{8}H to FeO would release approximately 1.1 $\times$ 10\textsuperscript{16} kg (\textasciitilde 4 hPa) of oxygen. This is~not enough for a complex biosphere. On the other hand, reducing ferric iron in Mars dust could induce a positive climate feedback, as it would make the surface darken, absorb more sunlight, and warm the climate: in a terraforming context, the red planet would turn green and black. 

Mars soil is~sulphate-rich, averaging 6.2 wt\% SO\textsubscript{3} 
\citep{yen2005,King2010}, and some rocks have 40 wt\% SO\textsubscript{3} \citep{franz2019}. Sulphate-reducing bacteria thrive in sediments on Earth that lack higher-energy electron acceptors; on~Mars, sulphate reduction could release CO\textsubscript{2} into the atmosphere for subsequent photosynthetic activity (e.g., \citealt{pedersen2013}). For example, consider the reaction of ferric sulphate into pyrite (FeS\textsubscript{2}) and iron oxide (eqs. \ref{eq:scheme10}-\ref{eq:scheme11}).  Assuming 6.2~wt\% average SO\textsubscript{3}, ferric sulphate reduction could support 3.1 $\times$ 10\textsuperscript{16} kg (\textasciitilde 11 hPa) of oxygen production from the soil, and an additional 7.2 $\times$ 10\textsuperscript{16} kg (\textasciitilde25 hPa) from dust adopting the~8.3~wt\%
SO\textsubscript{3} value from measurements by Curiosity/APXS \citep{berger2016}. Mars also has large, layered deposits rich in sulphates. \citet{hynek2019} estimates 4.8 $\times$ 10\textsuperscript{16} kg SO\textsubscript{3}-equivalent in the Meridiani sulphates alone. If the Meridiani sulphates were converted to pyrite, this would release 3.7~$\times$~10\textsuperscript{16} kg (\textasciitilde13 hPa) of oxygen. However, in contrast to soil processing which can happen naturally on civilisation-relevant timescales, the Meridiani deposits are rocks, and rock-to-soil conversion normally requires geologic timescales. Thus, making use of them would require breaking down the sulphate deposits through pedogenesis accelerated by large-scale artificial processing of the rock, before sulphate-reducing bacteria could begin to break down the sulphates contained within them. More sulphate deposits likely exist \citep{gaillard2009}, but may be too deeply buried for~reduction to pyrite to be practical. 

A problem for this biogeochemical oxygenation scheme is~that it needs very efficient oxidation of reduced organic matter. That is, the proposed geochemical network would need to bury, or store, reduced phases like FeO and FeS\textsubscript{2} without accompanying organic carbon. The needed recycling efficiency would depend on the final desired partial pressures of oxygen and carbon dioxide. Plausibly, each molecule of CO\textsubscript{2} would have to undergo many rounds of CO\textsubscript{2} fixation and anaerobic respiration prior to burial. On Earth since the Cambrian, burial of organic carbon and pyrite tends to co-occur \citep{bernerraiswell1983,raiswellberner1986}. In contrast, making a lot of atmospheric oxygen would require decoupling the burial of reduced sulphur and iron from the burial of organic carbon. We do not know if this is~possible. Building up  O\textsubscript{2} above the 20 hPa limit will require maintaining a careful balance between CO\textsubscript{2} in the atmosphere and sequestered organic matter that subsurface organisms could use for anaerobic respiration.

\begin{align}
\mathrm{\frac{1}{2}H_{2}O} & \rightarrow \mathrm{\frac{1}{4}O_{2}+H^{+}+e^{-}}\label{eq:scheme1}\\
\mathrm{2H_{2}O} & \rightarrow \mathrm{O_{2}+2H_{2}}\label{eq:scheme2}\\
\mathrm{CO_{2}+H_{2}O} & \rightarrow \mathrm{O_{2}+CH_{2}O}\label{eq:scheme3}\\
\mathrm{Ca(ClO_{4})_{2}+4CH_{2}O} & \rightarrow \mathrm{CaCl_{2}+4CO_{2}+4H_{2}O}\label{eq:scheme4}\\
\mathrm{Ca(ClO_{4})_{2}} & \rightarrow \mathrm{CaCl_{2}+4O_{2}}\label{eq:scheme5}\\
\mathrm{4Fe_{5}O_{8}H+5CH_{2}O} & \rightarrow \mathrm{5CO_{2}+20FeO+7H_{2}O}\label{eq:scheme6}\\
\mathrm{4Fe_{5}O_{8}H} & \rightarrow \mathrm{5O_{2}+20FeO+2H_{2}O}\label{eq:scheme7}\\
\mathrm{2Fe_{2}O_{3}+CH_{2}O} & \rightarrow \mathrm{4FeO+CO_{2}+H_{2}O}\label{eq:scheme8}\\
\mathrm{2Fe_{2}O_{3}} & \rightarrow \mathrm{4FeO+O_{2}}\label{eq:scheme9}\\
\mathrm{4Fe_{2}(SO_{4})_{3}+23CH_{2}O} & \rightarrow \mathrm{6FeS_{2}+2FeO+23H_{2}O+23CO_{2}}\label{eq:scheme10}\\
\mathrm{4Fe_{2}(SO_{4})_{3}} & \rightarrow \mathrm{23O_{2}+6FeS_{2}+2FeO}\label{eq:scheme11}
\end{align}

\begin{sidewaystable}
\includegraphics[width=1\textheight,totalheight=1\textwidth,keepaspectratio,page=1]{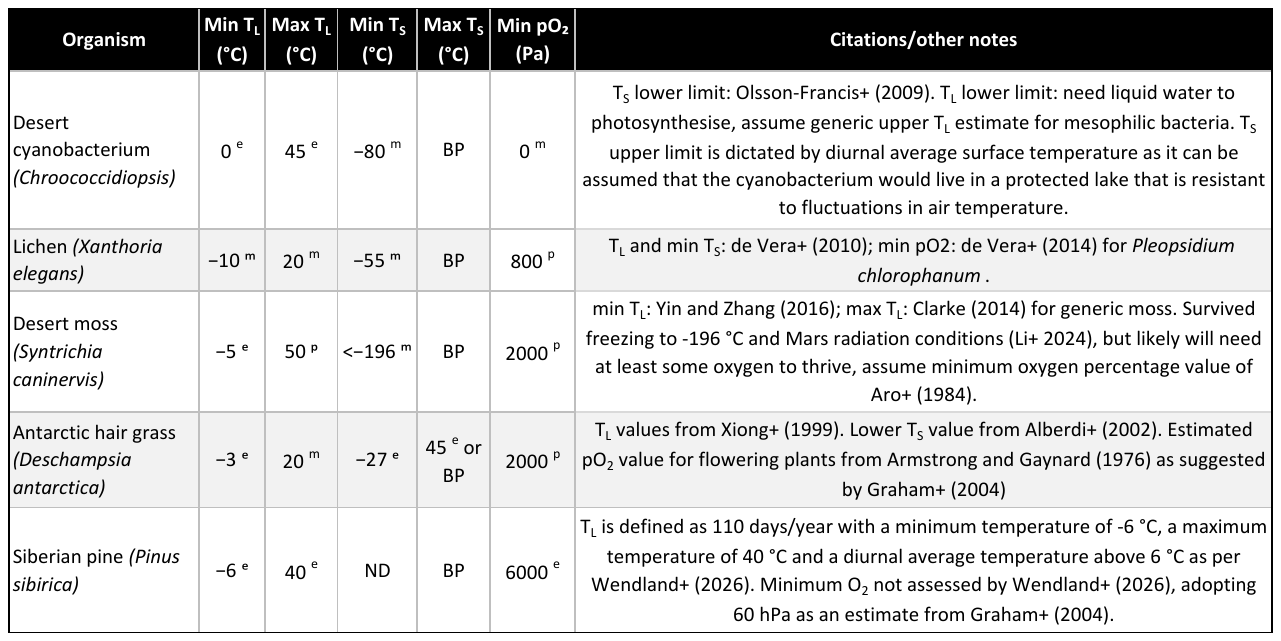}

\caption{Constraints on the survival of different pioneer organisms on~Mars,
defined by a) the temperature range within which the organism
can grow and metabolise for at least 60~sols (or other period of time
if specified) ($T_{L}$), b) lower and upper bounds on the temperatures
the organism can survive for at least one day ($T_{s}$) and c) minimum
oxygen partial pressure ($\mathrm{pO_{2}}$). BP = boiling point of water,
which is~a function of atmospheric pressure. ND = constraint not determined.
Each constraint is~either determined from experimentally
measured values in the literature (\emph{m}), values experimentally
measured for a similar proxy species (\emph{p}), or estimates (\emph{e}). }

\label{organismtable1}
\end{sidewaystable}
\begin{sidewaystable}
\includegraphics[width=1\textheight,totalheight=1\textwidth,keepaspectratio]{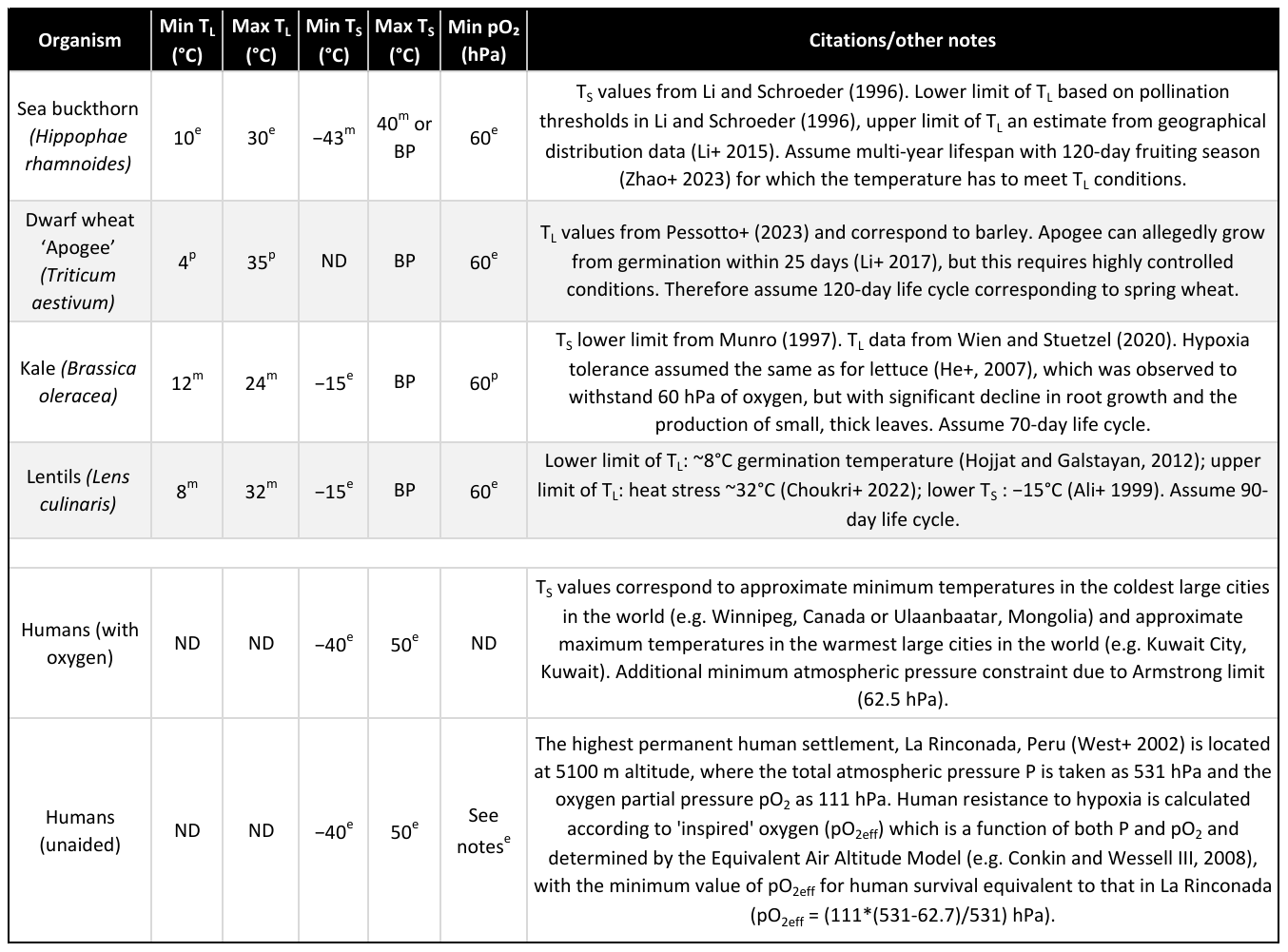}

\caption{As Table \ref{organismtable1}, but for humans and potential food
crops.}
\label{organismtables2}
\end{sidewaystable}

\begin{table}
\centering
\includegraphics[width=1\textwidth,keepaspectratio,trim=1cm 10cm 1cm 1cm, clip]{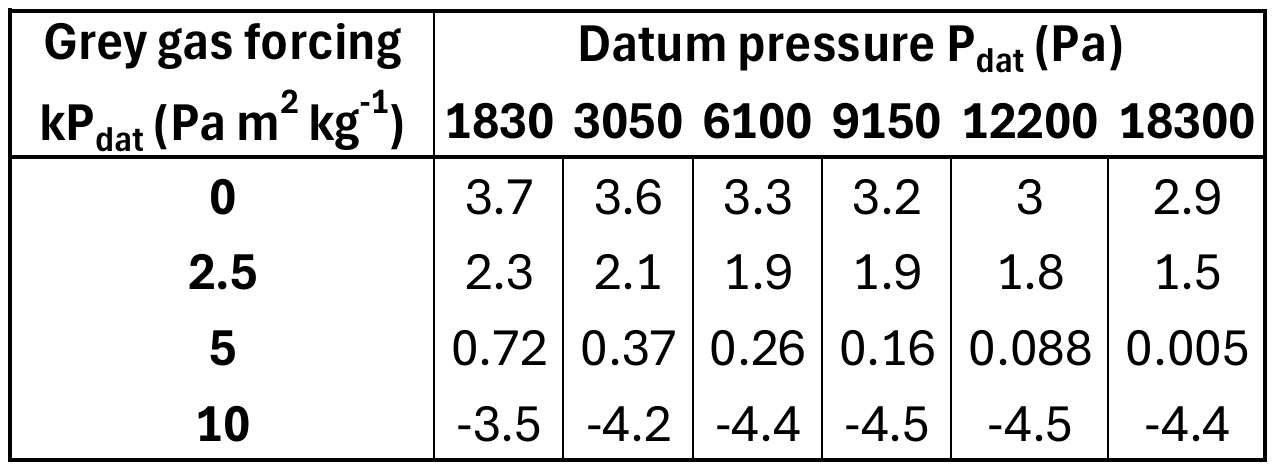}
\caption{Model error (energy imbalance). Net yearly average outgoing radiative flux (in W m\textsuperscript{-2}) at steady state at the top of atmosphere for each of the simulations presented in this work. Negative values indicate higher~incoming~flux than outgoing flux.}
\label{fluxleakstable}
\end{table}

\begin{figure}
\includegraphics[width=1\textwidth]{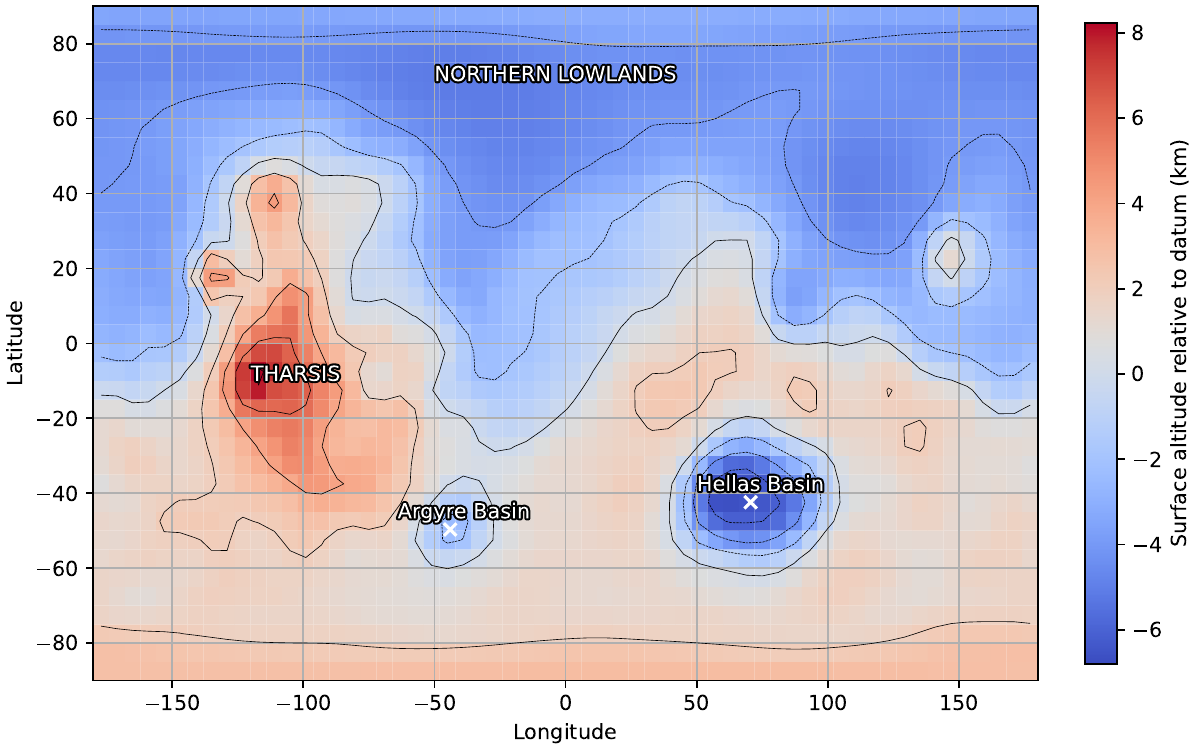}

\caption{Surface relief map of Mars, showing locations mentioned in the text.}
\label{topomap}
\end{figure}

\begin{sidewaysfigure}
\includegraphics[width=1\textwidth]{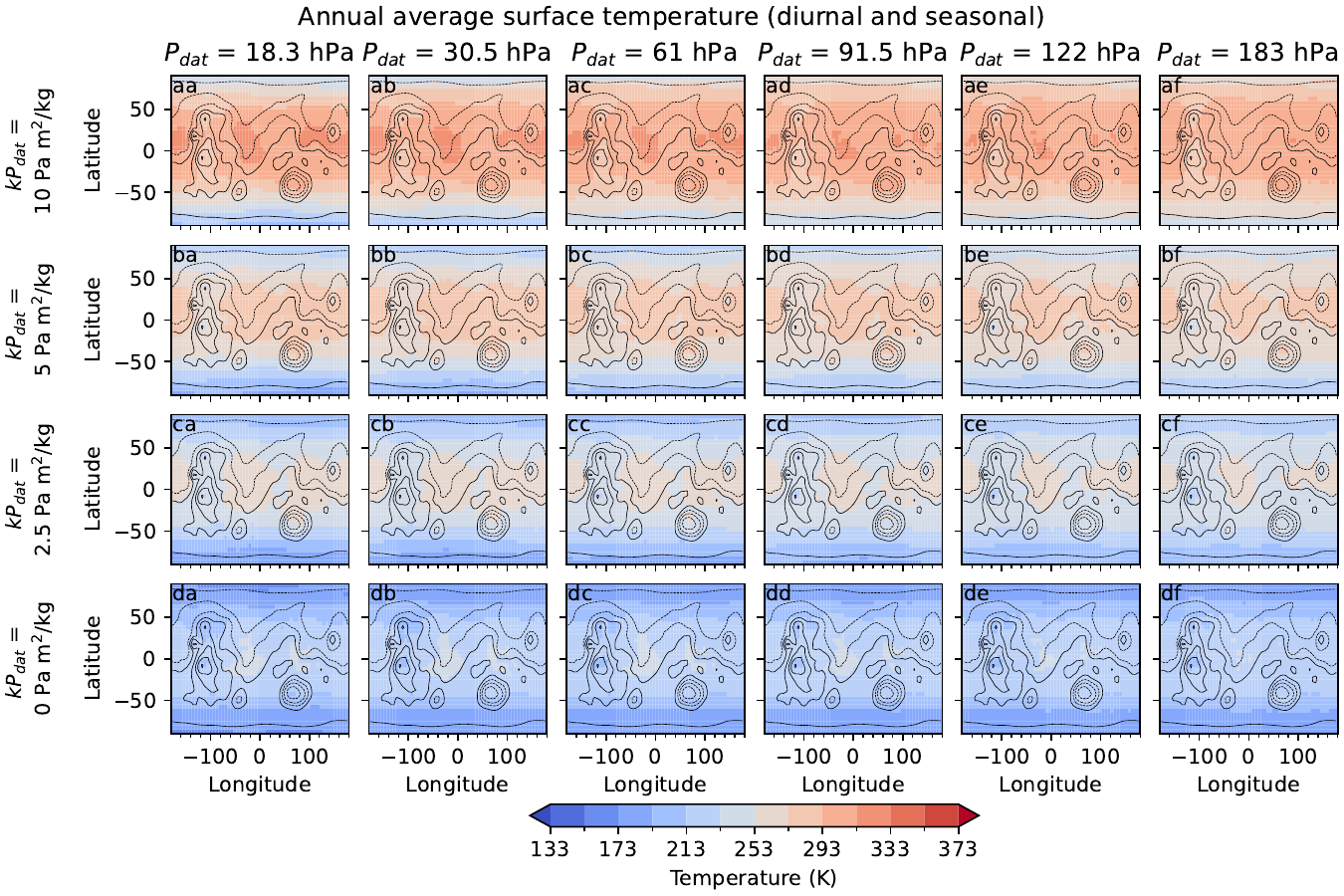}

\caption{Spatial variations in average surface temperatures as a function of
atmospheric pressure ($P_{dat}$) and grey gas forcing ($k P_{dat}$)}
\end{sidewaysfigure}
\begin{sidewaysfigure}
\includegraphics[width=1\textwidth]{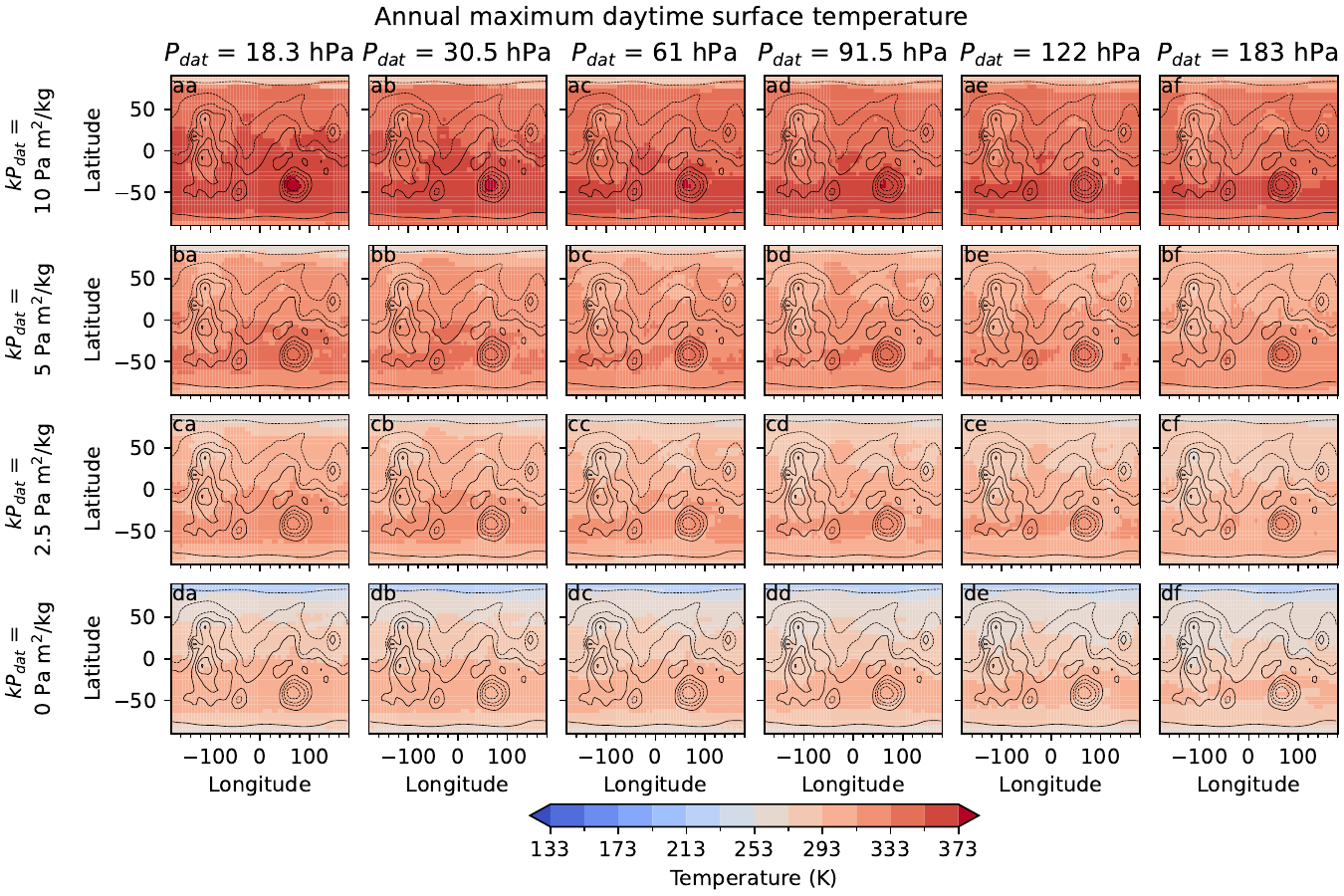}

\caption{Spatial variations in the maximum temperatures experienced over a
single Mars year, as a function of atmospheric pressure ($P_{dat}$)
and grey gas forcing ($k P_{dat}$).}

\label{maxtemps}
\end{sidewaysfigure}
\begin{sidewaysfigure}
\includegraphics[width=1\textwidth]{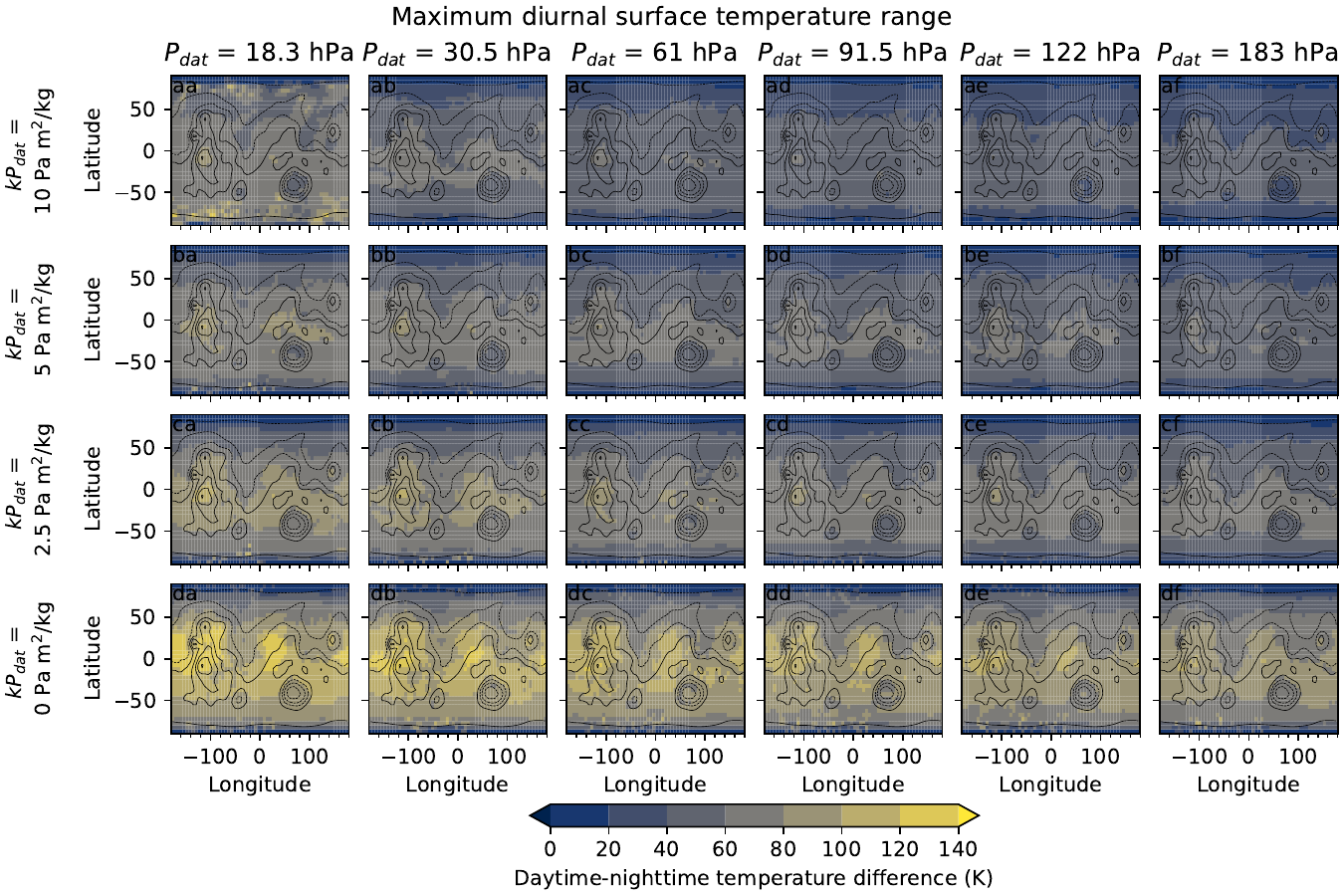}

\caption{Spatial variation in the maximum difference between the daytime high
(3~pm local time) and nighttime low temperature (3~am local time) experienced
for any day in a given~Mars~year, as a function of pressure ($P_{dat}$)
and grey gas forcing ($k P_{dat}$).}
\end{sidewaysfigure}

\begin{figure}
\includegraphics[width=1\textwidth]{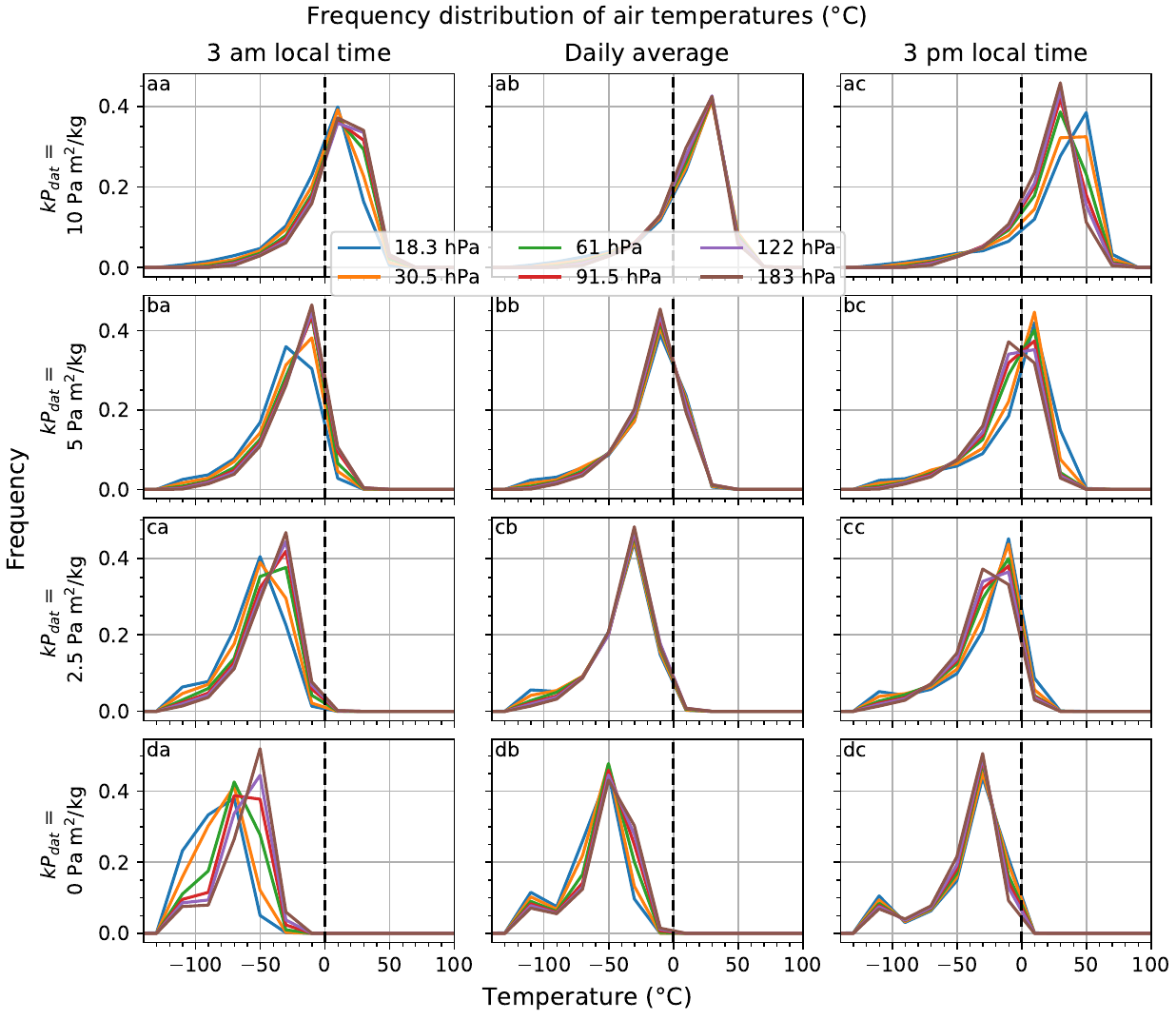}

\caption{Total frequency distribution of (left) nighttime, (centre) daily average and (right)~daytime~temperatures in the lowest layer of the atmosphere over all seasons, latitudes and longitudes for a single~Mars~year, as a function of pressure ($P_{dat}$)
and grey gas forcing ($k P_{dat}$).}
\end{figure}

\begin{figure}
\includegraphics[width=1\textwidth]{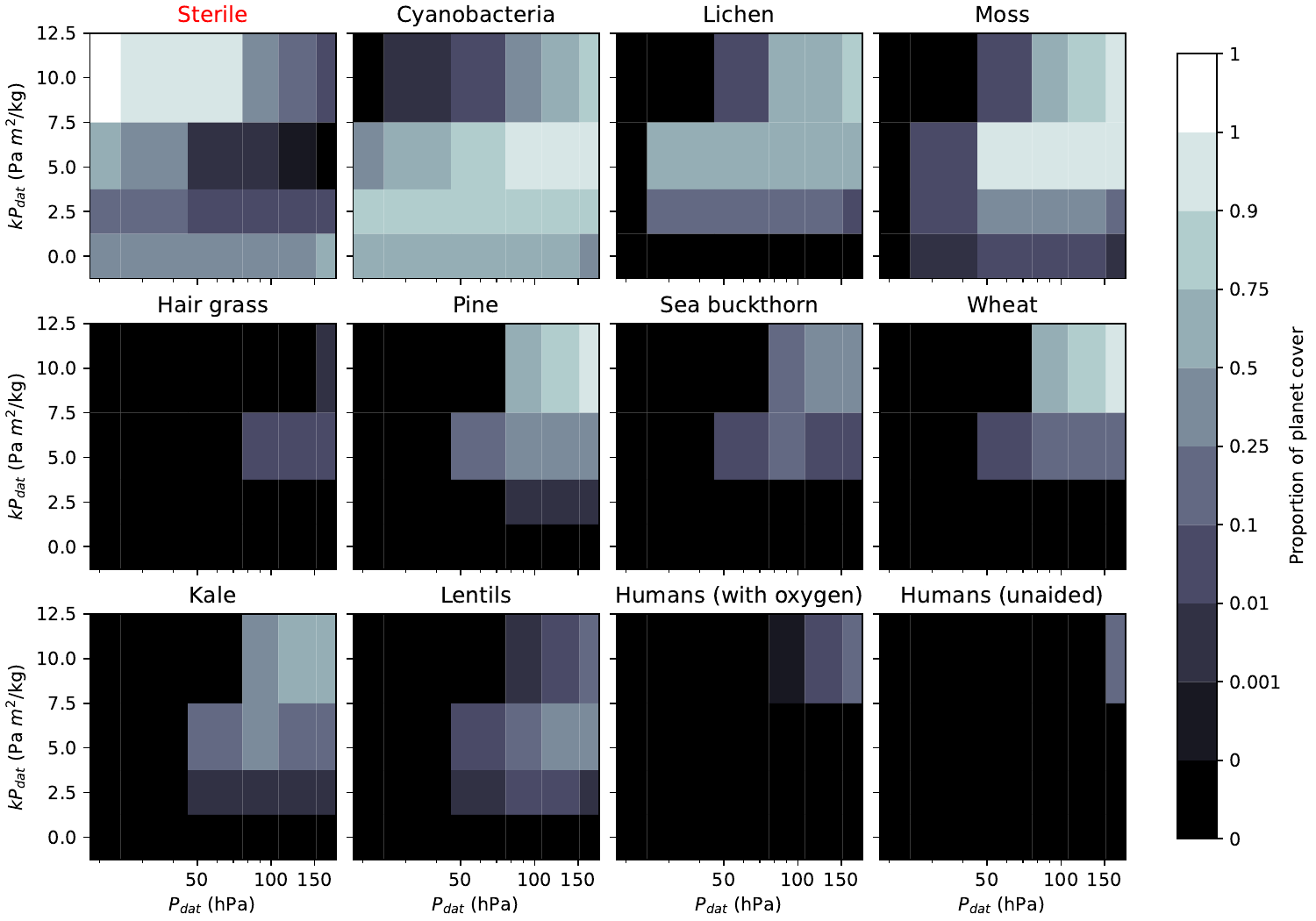}

\caption{Fraction of the Mars surface in which the temperature and oxygen environment
permits each of~the~listed organisms, as a function of pressure ($P_{dat}$)
and grey gas forcing ($k P_{dat}$). The top left panel indicates
the~proportion of the surface on which none of the listed organisms
can live.}
\label{ecotables}
\end{figure}

\begin{figure}
\includegraphics[width=1\textwidth]{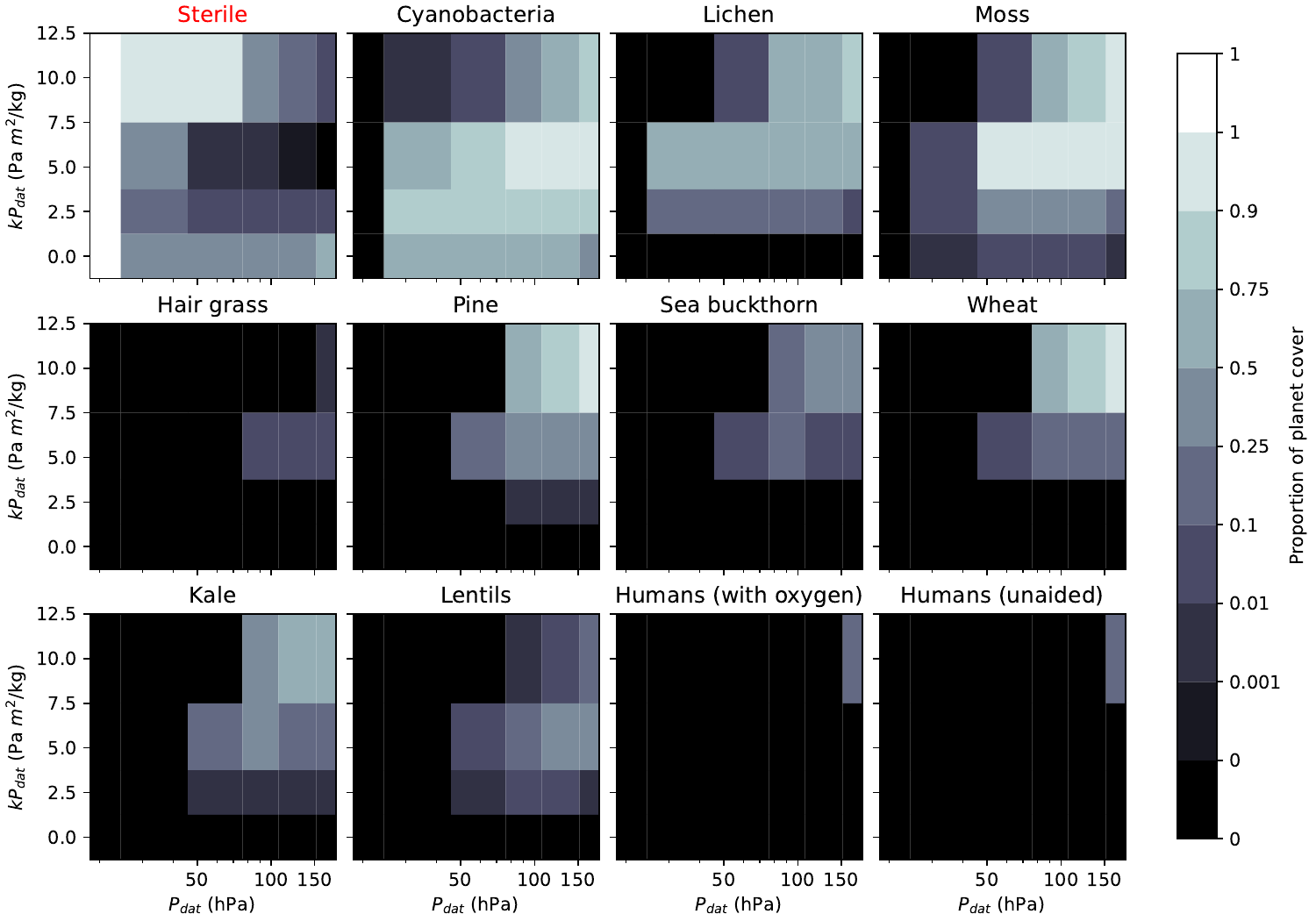}

\caption{As Fig. \ref{ecotables}, but with an additional maximum constraint
of 5 mSv of galactic cosmic radiation for human survivability and
0.0025 W m\protect\textsuperscript{-2} of UVC radiation for the survivability
of cyanobacteria.}

\label{ecotables-1}
\end{figure}
\begin{sidewaysfigure}
\includegraphics[width=1\textwidth]{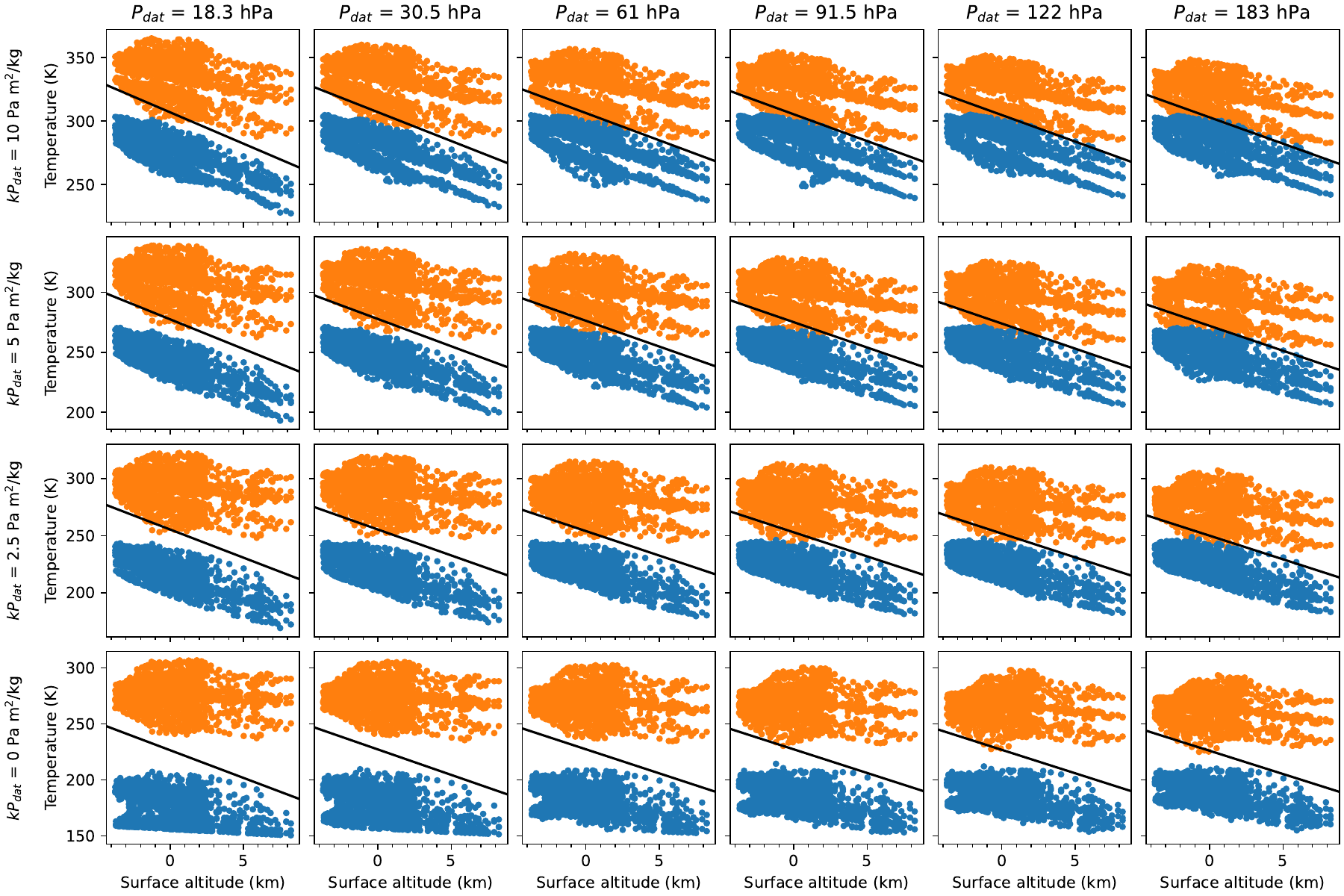}

\caption{The lapse rate in surface temperature is~more sensitive to greenhouse warming than it is~to atmospheric pressure (see \citealt{fan2023} for details and theoretical discussion). Spread of (blue) nighttime and (orange) daytime surface temperatures
within 20° latitude of the Equator, as a function of surface
altitude. An isothermal atmosphere does not vary in temperature with altitude, while an atmosphere in radiative-convective
equilibrium parallels the dry
adiabatic lapse rate (solid black line for reference). Increased grey gas
forcing results in a larger increase in the atmospheric temperature
gradient than increased atmospheric pressure. For moderate grey gas opacities, nighttime
temperatures approximate a dry adiabat more closely than do daytime temperatures.}
\label{lapserate}
\end{sidewaysfigure}
\begin{figure}
\includegraphics[width=1\textwidth,page=1]{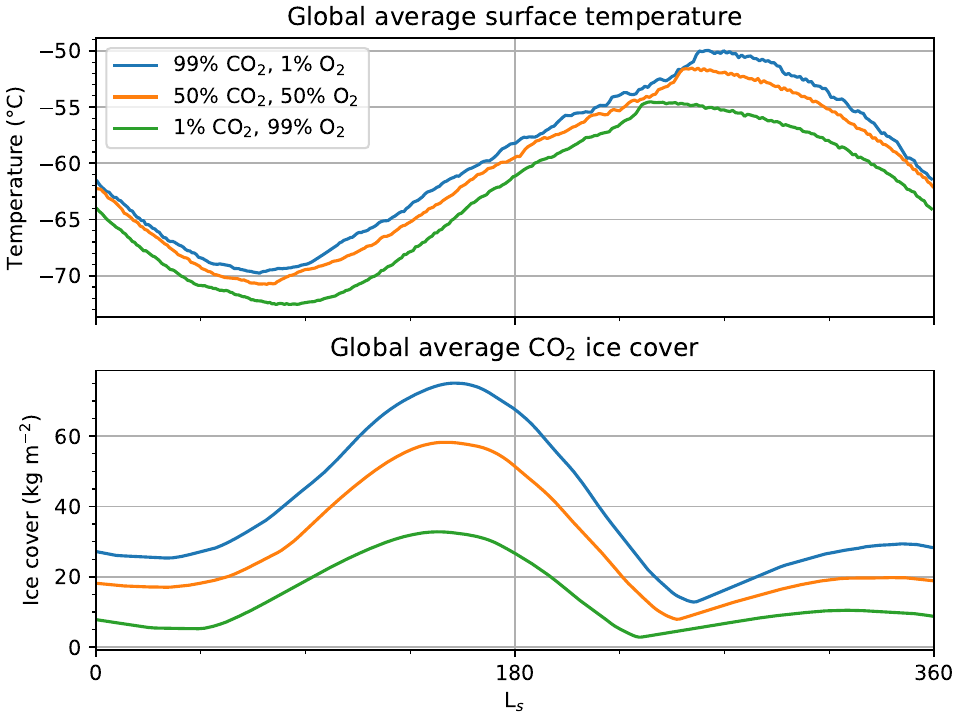}
\caption{(top) Sensitivity of global average surface temperature to the volume
mixing ratios of CO\protect\textsubscript{2} to~O\protect\textsubscript{2}
in the atmosphere, with constant $P_{dat}=122$ hPa and $k P_{dat}=0$.
Although a pure CO\protect\textsubscript{2} atmosphere is~only~1~K warmer on average than an equal mix of CO\protect\textsubscript{2}-O\protect\textsubscript{2},
a pure O\protect\textsubscript{2} atmosphere leads to a \textasciitilde5 K decrease
in global~average~temperatures relative to a pure CO\protect\textsubscript{2}
atmosphere. Seasonal lag in average surface temperatures also~correlates with surface CO\protect\textsubscript{2} ice condensation (bottom panel).}

\label{co2o2mix}
\end{figure}
\nocite{clarke2014,olssonfrancis2009}

\nocite{devera2010,devera2014}

\nocite{Yinzhang2016,Li2024,aro1984}

\nocite{Xiong1999,alberdi2002,armstronggaynard1976,graham2004}

\nocite{wendland2026}

\nocite{Lischroeder1996,zhao2023,li2015evaluation}

\nocite{li2017wheat,pessotto2023}

\nocite{munro1997,wien2020,he2007}

\nocite{hojjat2012effects,Choukri2022,ALI_JOHNSON_STUSHNOFF_1999}

\nocite{west2002larinconada}